\documentclass[final,5p,times,twocolumn,authoryear]{elsarticle}
\usepackage{amsmath,xcolor}
\usepackage{graphicx}
\usepackage{epstopdf}
\usepackage{url}
\usepackage{amsthm}
\usepackage{comment}
\usepackage{amssymb}
\usepackage{caption}
\usepackage{subcaption}
\usepackage{amsfonts}
\usepackage[ruled,norelsize,linesnumbered]{algorithm2e}
\usepackage{soul,xcolor}
\usepackage{bm}
\usepackage{mathtools}
\usepackage{tabularx}
\usepackage{float}
\usepackage{array}
\usepackage{picins}
\usepackage{threeparttable}
\makeatletter
\newcommand{\removelatexerror}{\let\@latex@error\@gobble}
\makeatother
\biboptions{sort&compress}
\journal{Automatica}
\newtheorem{theorem}{Theorem}
\newtheorem{lemma}{Lemma}
\newtheorem{remark}{Remark}
\newtheorem{definition}{Definition}
\newtheorem{assumption}{Assumption}

\makeatletter
\def\hlinewd#1{%
	\noalign{\ifnum0=`}\fi\hrule \@height #1 %
	\futurelet\reserved@a\@xhline}
\makeatother
\removelatexerror
\newcolumntype{L}[1]{>{\raggedright\let\newline\\\arraybackslash\hspace{0pt}}m{#1}}
\newcolumntype{C}[1]{>{\centering\let\newline\\\arraybackslash\hspace{0pt}}m{#1}}
\newcolumntype{R}[1]{>{\raggedleft\let\newline\\\arraybackslash\hspace{0pt}}m{#1}}
\begin{document}
	
	\begin{frontmatter}
		
		\title{Privacy-Preserving Nonlinear Cloud-based Model Predictive Control via Affine Masking\tnoteref{footnoteinfo}}		
		\tnotetext[footnoteinfo]{This work is supported by National Science Foundation grant \#2045436. The material in this paper was not presented at any conference.}
		
		\author[MSU]{Kaixiang Zhang}\ead{zhangk64@msu.edu}
		\author[MSU]{Zhaojian Li\corref{cor1}}\ead{lizhaoj1@egr.msu.edu}
		\cortext[cor1]{Corresponding author.}
		\author[Clemson]{Yongqiang Wang}\ead{yongqiw@clemson.edu}
		%\author[UMICH]{Nan Li}\ead{nanli@umich.edu}
		\author[AU]{Nan Li}\ead{nanli@auburn.edu}            % e-mail address

		\address[MSU]{Department of Mechanical Engineering, Michigan State University, East Lansing, MI 48824, USA.}
		\address[Clemson]{Department of Electrical and Computer Engineering, Clemson University, Clemson, SC 29634, USA.}
		%\address[UMICH]{Department of Aerospace Engineering, University of Michigan, Ann Arbor, MI 48109, USA.}
		\address[AU]{Department of Aerospace Engineering, Auburn University, Auburn, AL 36849, USA.}
	
	\begin{abstract}
		With the advent of 5G technology that presents enhanced communication reliability and ultra low latency, there is renewed interest in employing cloud computing to perform high performance but computationally expensive control schemes like nonlinear model predictive control (MPC). Such a cloud-based control scheme, however, requires data sharing between the plant (agent) and the cloud, which raises privacy concerns. This is because privacy-sensitive information such as system states and control inputs has to be sent to/from the cloud and thus can be leaked to attackers for various malicious activities.  % Data sharing in cloud-based control is ubiquitous, which raises the concern that the private information of plant could be leaked when the cloud is malicious or communication channels are wiretapped.
		In this paper, we develop a simple yet effective affine masking strategy for privacy-preserving nonlinear MPC. Specifically, we consider external eavesdroppers or honest-but-curious cloud servers that wiretap the communication channel and intend to infer the plant's information including state information and control inputs. 
		%An affine transformation mechanism is designed to enable privacy preservation without affecting the MPC calculation.
		An affine transformation-based privacy-preservation mechanism is designed to mask the true states and control signals while reformulating the original MPC problem into a different but equivalent form. We show that the proposed privacy scheme does not affect the MPC performance and it preserves the privacy of the plant such that the eavesdropper is unable to identify the actual value or even estimate a rough range of the private state and input signals.
		%such that the eavesdropper cannot discover the private state and input signals with any guaranteed accuracy. 
		The proposed method is further extended to achieve privacy preservation in cloud-based output-feedback MPC. Simulations are performed to demonstrate the efficacy of the developed approaches.
	\end{abstract}
\end{frontmatter}

\thispagestyle{empty}
\pagestyle{empty}
\setlength{\abovecaptionskip}{0pt}
\setlength{\belowcaptionskip}{4pt}
\setlength{\textfloatsep}{0pt}

%\vspace{-10 pt}
\section{Introduction} 
%\vspace{-5 pt}
Model predictive control (MPC) is an optimal control paradigm that can explicitly handle system constraints and has enjoyed great successes over the past decade \citep{MayneAUTO2014,Li_stochastic,AllenspachAUTO2021,LiuTIE2016}. %as .  extensively studied for the control of constrained systems and is emerging as an effective tool for different applications \cite{MayneAUTO2014,Li_stochastic,AllenspachAUTO2021,LiuTIE2016}. 
Despite their outstanding performances, conventional MPC implementations involve solving an online optimization problem that requires substantial computation power, especially for nonlinear and complex systems. This hinders the deployment of MPC in many resource-limited cyber-physical systems with real-time constraints such as autonomous vehicles and mobile robots. Cloud-based MPC -- outsourcing the heavy computation to the cloud with superior computational resources -- has received renewed attention \citep{li2021cloudassisted,NilsCDC2020,SultangazinTAC2021}, partly attributed to the advancement in 5G technologies that can provide reliable communication with negligible latency. %  typically requires Conventional MPC-based control systems usually embed all components, including sensor, controller and actuator, into a local unit and need to solve an open-loop optimal control problem at each sampling time. This practice leads to a high cost of onboard computation, complex installation and lack of flexibility, hindering the application of MPC in resource-constrained control systems, e.g., connected vehicles and micro robots. To address the resource-constrained issue, a possible way is to integrate cloud computing technologies into the control systems.   

%Cloud computing is a computing paradigm that has evolved significantly over the past few years. 
In brief, cloud computing is a unified platform that  provides on-demand computing power and data storage services to users \citep{grossman2009case}. The cloud can offer superior computational capabilities to execute advanced (and computationally expensive) control strategies like nonlinear MPC, as well as incorporate real-time crowdsourced information as a preview to increase situational awareness and enhance system performance
\citep{ZL_suspension,comfort,Li_Safe_Journal,fuel_economy1}.  A general setup for cloud-based MPC is as follows. First, the plant sends the measured (or estimated) states to the cloud. The cloud then solves a pre-specified MPC problem and sends back the optimal control actions. The system evolves over one step and the process is then repeated. The aforementioned setup has several advantages, including high performance (if the communication has negligible latency), easy deployment, and convenient  modification when needed, among others. However, the system states/measurements and control actions need to be transmitted between the cloud and the local agent, raising concerns that outsourcing computation to a cloud might leak private information (e.g., sensor measurements and system states) to an eavesdropper or an untrusted cloud. In fact, several studies have shown that exposing local agent's information to connectivity can indeed lead to security vulnerabilities and various malicious activities \citep{PetitTITS2015,Munteanu2018,Xu2021IS}. 

Considering the aforementioned concerns and the growing awareness of security in cyber-physical systems, it is imperative to protect the privacy of agents if cloud-based control is used. As such, several privacy preservation schemes for cloud-based MPC have been proposed, which can be mainly categorized into homomorphic encryption-based methods \citep{NilsCDC2020,SchulzeCSL2018,AlexandruCDC2018,DarupIFAC2018} and algebraic transformation based methods \citep{Xu2015ACM,Xu2017,SultangazinTAC2021,Naseri2022ECC}. The homomorphic encryption-based methods exploit cryptography to mask privacy-sensitive information (e.g., system states) while still enabling the cloud to perform the MPC computation with encrypted data. In \cite{SchulzeCSL2018}, homomorphic encryption is used to design a secure explicit MPC scheme for linear systems with state and input constraints. The encrypted fast gradient method and proximal gradient method are developed in \cite{AlexandruCDC2018} and \cite{DarupIFAC2018}, respectively, to achieve implicit MPC for linear systems with input constraints. Despite strong privacy guarantees for the cloud-based MPC, the induced encryption and decryption procedures are quite computationally heavy, which is thus not suitable for systems with limited onboard resources and stringent real-time constraints. 

Different from the homomorphic encryption-based methods, the algebraic transformation-based approaches rely on introducing transformation maps that act as masks, rendering the real signals of a local agent indiscernible by the attacker. More specifically, the main idea of the algebraic transformation methods is to design appropriate transformation maps to protect privacy-sensitive signals and construct a different but equivalent MPC problem. Without knowing the original MPC problem, the cloud will solve the equivalent MPC problem and provide the plant with the corresponding optimal control action. By using inverse transformation maps, the plant can recover the optimal control action to the original problem. This idea has been initially applied to accomplish privacy preservation in optimization \citep{Weeraddana2013CDC,Weeraddana2017IFAC,Mangasarian2011,WangINFOCOM2011} and then extended to cloud-based MPCs. For example, in \cite{Xu2015ACM}, non-singular matrices are utilized to produce a transformation mechanism for linear MPC in networked control system. In \cite{Xu2017}, orthogonal matrices are combined with homomorphic encryption to design a hybrid privacy preservation scheme for output-feedback MPC. In \cite{Naseri2022ECC}, random transformations are utilized to achieve privacy preservation for set-theoretic MPC. Furthermore, isomorphisms and symmetries are adopted in \cite{SultangazinTAC2021} as a source of transformation to protect the privacy of system signals.

In this paper, a privacy-preserving cloud-based nonlinear MPC framework is developed to protect system privacy (e.g., states, inputs) via an affine transformation scheme (which is a form of algebraic transformation). We first show that if the cloud is an honest-but-curious adversary or there exists an external eavesdropper, the conventional cloud-based MPC architecture cannot protect the private information of the plant. An affine transformation-based privacy mechanism is then designed to mask the real system state and input signals. With the affine transformation, we reformulate the original MPC problem into a different but equivalent one, which is solved by the cloud.  Solution to the equivalent MPC problem is then received by the local agent and transformed via simple inverse affine transformation to recover the solution to the original problem. 
A privacy definition is introduced to show that the proposed affine transformation scheme can protect the private system state and input signals from being inferred by the attacker. 

The major contributions of this paper include the following. First, we develop a privacy-preserving cloud-based  MPC for a class of nonlinear systems. While studies on privacy-preserving cloud MPC for linear systems exist (see e.g., \cite{NilsCDC2020,SchulzeCSL2018,AlexandruCDC2018,DarupIFAC2018,Xu2015ACM,Xu2017,SultangazinTAC2021,Naseri2022ECC}), to the authors' best knowledge, this is the first work on privacy-preserving cloud MPC for a class of nonlinear systems with general constraints. Using cloud computing for nonlinear and complex systems makes most practical sense as recent advances in compact and powerful onboard computation units are enabling real-time implementations for linear MPCs (but still not for nonlinear MPCs) \citep{GM-MPC}. We mask the privacy-sensitive signals via affine transformation and reformulate a compatible nonlinear MPC that is equivalent to the original problem, thus with no performance degradation. 
%Different from the homomorphic encryption methods \cite{AlexandruCDC2018,DarupIFAC2018} that are designed for linear MPC with particular forms, the proposed affine transformation method can work for a wide range of nonlinear MPC problems. 
Furthermore, the affine transformation method is light-weight in computation, which makes it easily applicable to cloud-based control. Second, a new privacy definition, $\infty$-diversity with unbounded diameter, is introduced that is suitable for the considered real-time cyber-physical systems. Third, we extend the developed framework to cloud-based nonlinear output-feedback MPC to achieve privacy preservation for nonlinear systems with only output feedback. Finally, simulation examples are presented to demonstrate the efficacy of the developed framework.
%The proposed approach is inspired by the algebraic transformation based works designed for linear systems \citep{Xu2015ACM,Xu2017,SultangazinTAC2021}, but there exist several differences. The scheme proposed in \cite{Xu2015ACM} only works for special objective functions and linear input constraints, and neither state nor input constraints are well considered in \cite{Xu2017}. Instead, our developed approach can be applied to more general MPC problems as we consider nonlinear systems, objective function described by general quadratic form, and state and input constraints. 
%The recent work \cite{SultangazinTAC2021} proposes a novel transformation scheme based on isomorphisms and symmetries, which quantifies the guaranteed privacy via the dimension of the set that describes the uncertainty experienced by the adversary. 
%We use a similar affine transformation mechanism and communication protocol as presented in \cite{SultangazinTAC2021} to conceal the sensitive information.
%Different from the work \cite{SultangazinTAC2021} that quantifies the privacy via the dimension of the manifold that describes the uncertainty experienced by the adversary, we use the set cardinality and diameter to define the privacy notion for cloud-based nonlinear MPC. 
%tNote that the set dimension based privacy quantification in \cite{SultangazinTAC2021} is derived based on the characteristics of linear systems, which cannot be directly applied to nonlinear systems.	
The proposed approach draws inspiration from algebraic transformation-based methods developed for linear systems \citep{Xu2015ACM,Xu2017,SultangazinTAC2021}, but there exist significant differences between our work and these references. The scheme proposed in \cite{Xu2015ACM} is limited to special objective functions and linear input constraints, and in \cite{Xu2017}, neither state nor input constraints are considered. In contrast, our approach is designed to address more general MPC problems, encompassing nonlinear systems, objective functions described by general quadratic form, and accounting for state and input constraints.	
To conceal sensitive information, we employ an affine transformation mechanism and communication protocol similar to that presented in \cite{SultangazinTAC2021}. However, different from the work of \cite{SultangazinTAC2021} which quantifies privacy via the dimension of the manifold that describes the diversity experienced by the adversary,
%unlike the privacy quantification approach in \cite{SultangazinTAC2021}, which measures privacy based on the dimension of the manifold describing uncertainty experienced by the adversary, 
we tailor the privacy notion for cloud-based nonlinear state-feedback and output-feedback MPC by using set cardinality and diameter. Note that the set dimension-based privacy quantification in \cite{SultangazinTAC2021} is derived based on the state/input/output matrices of linear systems and cannot be directly applied to nonlinear systems. Our privacy notion works for general nonlinear systems, and it requires that after observing the released data, the adversary has infinite uncertainties on each of its interested entries and the difference between the possible uncertainties could be arbitrarily large.  

%In addition, we also extend the developed method to achieve the privacy preservation in output-feedback MPC over the cloud. Simulation results are given to validate the performance of the affine transformation scheme. 
%There are important differences between our work and the existing studies. The works \cite{SchulzeCSL2018,AlexandruCDC2018,DarupIFAC2018,Xu2015ACM,Xu2017,SultangazinTAC2021} consider the MPC or optimal control of linear systems, while in this work, we focus on privacy protection of state-feedback and output-feedback MPC for nonlinear plants. Different from the homomorphic encryption methods \cite{AlexandruCDC2018,DarupIFAC2018} that are designed for particular MPC solutions, the proposed affine transformation method can work for more general MPC problem. Furthermore, the affine transformation method is light-weight in computation, which makes it easily applicable to cloud-based control.

The rest of this paper is organized as follows. Section~\uppercase\expandafter{2} introduces the problem formulation including cloud-based MPC and the attack model. Section~\uppercase\expandafter{3} presents the developed privacy preservation scheme via affine transformations. We then extend the scheme for output-feedback MPC in Section~\uppercase\expandafter{4}. Simulations are presented in Section~\uppercase\expandafter{5}, and finally Section \uppercase\expandafter{6} concludes this paper.

%\vspace{-3mm}
\section{Problem Formulation} \label{sec_preliminary}
In this section, we present relevant background of the considered privacy-preserving cloud-based MPC problem. Specifically, we first introduce the conventional cloud-based MPC framework with no privacy protection, followed by a description of the privacy attack model considered in this paper. 

%\vspace{-3mm}
\subsection{Cloud-based MPC} \label{subsec_cloud MPC}
%Each CE-NCS has two layers: cyber layer and physical layer. The cyber layer consists of wireless communications and a cloud, while the physical layer incorporates a plant, actuators, and sensors, and a controller. The integration of the controller with the cloud constitutes a cloud-based controller. Fig. \ref{} illustrates a feedback architecture of a CE-NCS. The controller of a CE-NCS is structurally different from other general NCSs. Instead of solving an off-line control problem, the controller of a CE-NCS formulates a dynamic optimization problem based on the sensor data, and out- sources the computations of the control decisions to a cloud. After receiving the solutions from the cloud, the controller sends them to the actuators of the physical system. 

%To describe in detail the architecture of a CE-NCS, we first introduce the physical layer control problem. In this work, we present a discrete-time linear system to capture the dynamics of the physical plant and use Model Predictive Control (MPC) framework to de- sign optimal control to stabilize the system. The computations of the MPC control inputs are outsourced to the cloud, which can be subject to adversarial attacks.

We consider a class of nonlinear systems which can be described by the following control-affine discrete-time model:
\begin{equation} \label{equ_system}
	x(k+1) = \Phi(x(k), u(k)) = f(x(k)) + g(x(k))u(k),
\end{equation}
where $x(k) \in \mathbb{R}^{n}$ is the system state, $u(k) \in \mathbb{R}^{m}$ is the control input, $\Phi(\cdot, \cdot) \in \mathbb{R}^{n}$, $f(\cdot) \in \mathbb{R}^{n}$ and $g(\cdot) \in \mathbb{R}^{n\times m}$ are nonlinear continuous functions characterizing the system dynamics. At each sampling instant $k$, the following nonlinear MPC problem is solved: 
\begin{equation} \label{equ_MPC}
	%{\small	
		\begin{aligned}
			\textbf{P-1}:\;\; &\min_{\bm{U}_{k}} J_{N}(x(k),\bm{U}_{k}) 
			\\
			&= \sum_{i=0}^{N-1}(x_{i|k}^{\top}Qx_{i|k} + q^{\top}x_{i|k} + u_{i|k}^{\top}Ru_{i|k} + r^{\top}u_{i|k})
			\\
			&\;\;\;\; + x_{N|k}^{\top}Q_{f}x_{N|k} + q_{f}^{\top}x_{N|k},
			\\
			\text{s.t.}\;\; & x_{i+1|k} = f(x_{i|k})+g(x_{i|k})u_{i|k}, i=0, \cdots, N-1,
			\\
			& x_{i|k} \in \mathcal{X}, i=1, \cdots, N-1,
			\\
			& u_{i|k} \in \mathcal{U}, i=0, \cdots, N-1,
			\\
			& x_{N|k} \in \mathcal{X}_{f},
			\\
			& x_{0|k} = x(k), \bm{U}_{k}=\begin{bmatrix}
				u_{0|k}^{\top}, \cdots, u_{N-1|k}^{\top}
			\end{bmatrix}^{\top},
		\end{aligned}
		%}
\end{equation}
which is a receding horizon optimal control problem with state and input constraints. In \eqref{equ_MPC}, $J_{N}(\cdot, \cdot) \in \mathbb{R}$ is the cost function with $Q \in \mathbb{R}^{n\times n}$, $q\in \mathbb{R}^{n}$, $R\in \mathbb{R}^{m\times m}$, $r\in \mathbb{R}^{m}$, $Q_{f} \in \mathbb{R}^{n\times n}$ and $q_{f} \in \mathbb{R}^{n}$ being weighting matrices and vectors; $x_{i|k}$ and $u_{i|k}$ are, respectively, the predicted system state and the input $i$ time steps ahead of current time instant $k$; $N \in \mathbb{N}_{+}$ is the prediction horizon; $\mathcal{X} \subset \mathbb{R}^{n}$ and $\mathcal{U} \subset \mathbb{R}^{m}$ are state and input constraint sets, respectively, and $\mathcal{X}_{f} \subset \mathbb{R}^{n}$ is the terminal set.  

In a conventional MPC, the optimization problem \eqref{equ_MPC} is solved at each time step based on the current state $x(k)$, and the first element of the optimal input sequence $\bm{U}^{*}_{k}= {\small\begin{bmatrix}
		u_{0|k}^{*\top}, \cdots, u_{N-1|k}^{*\top}
	\end{bmatrix}^{\top}}$ is applied to the system, i.e., $u(k)=u_{0|k}^{*}$, and the system evolves over one step. The process is then repeated. With gentle assumptions and by appropriately selecting the weighting matrix $Q_{f}$ and terminal set $\mathcal{X}_{f}$, the resulting closed-loop system can achieve guaranteed recursive feasibility and asymptotical stability \citep{Rawlings2017MPC}.
\begin{figure}[!t]
	\centering
	\hspace{0 in}
	\includegraphics[width=2.9 in]{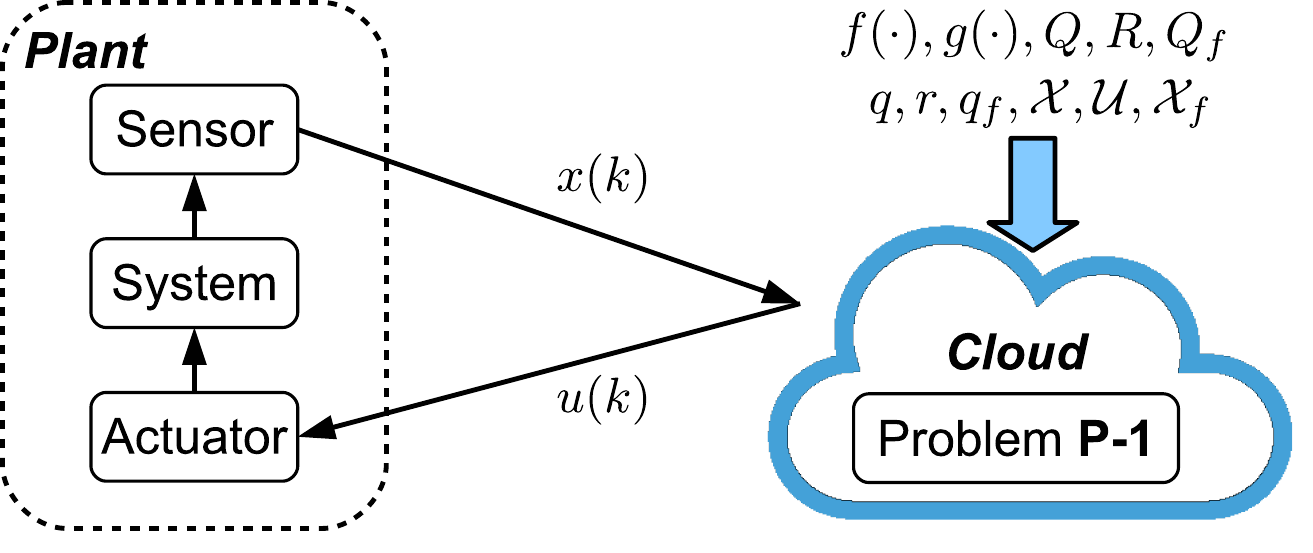}
	\vspace{1mm}
	\caption{Cloud-based conventional MPC architecture.}
	\label{fig_unsecure MPC}
	%\vspace{-5mm}
\end{figure}

The optimization problem in \eqref{equ_MPC} is a nonlinear programming problem that requires significant computation power, which is very challenging to solve onboard considering limited onboard computation and stringent real-time constraints for many cyber-physical systems. 
This challenge is exacerbated when the dimension of the system state and the prediction horizon are large. To address this problem, cloud-based MPC is a viable framework where the complex computation is outsourced to the cloud that has superior computational power. The ultra low latency brought by 5G technologies makes this framework especially appealing. Specifically,  the common cloud-based MPC architecture is shown in Fig.~\ref{fig_unsecure MPC}, which includes the following two phases:
\begin{itemize}
	\item Handshaking Phase: The plant sends 
	\[
	\left\lbrace f(\cdot), g(\cdot), Q, q, R, r, Q_{f}, q_{f}, \mathcal{X}, \mathcal{U}, \mathcal{X}_{f} \right\rbrace
	\]
	to the cloud, that is, the necessary information for the cloud to set up the nonlinear programming problem in \eqref{equ_MPC}.
	
	\item Execution Phase: At each time step $k$, the plant first sends its state $x(k)$ to the cloud. Then the cloud computes $u(k)$ by solving the optimization problem \eqref{equ_MPC} and sends the resultant $u(k)$ to the plant. Finally, the plant applies $u(k)$ to the actuators and the system evolves over one step.
\end{itemize} 

%\vspace{-3mm}
\subsection{Attack Model}  
As described above, for the conventional cloud-based MPC, the plant needs to provide the cloud with the system state, dynamic model, objective function, and constraints, which may contain confidential information that needs to be protected from an external eavesdropper or the untrusted cloud. In this paper, we consider the following two attack models:
\begin{itemize}
	\item \emph{Eavesdropping attacks} are attacks in which an external eavesdropper wiretaps communication channels to intercept exchanged messages in an attempt to learn the information about sending parties.
	\item \emph{Honest-but-curious attacks} are attacks in which the untrusted cloud follows all protocol steps correctly but is curious and collects all received intermediate data in an attempt to learn the information about the plant.
\end{itemize}

%that the cloud is an \textit{honest-but-curious adversary}: The cloud commits to following all protocol steps for the computations subscribed by the plant but will try to extract and leak privacy-sensitive information by recording all communicated messages.
In particular, we consider the case that the privacy-sensitive information is contained in the system state $x(k)$ and input $u(k)$. 
When cloud-based MPC is adopted in some specific areas, such as intelligent vehicle and smart grid, the disclosure of the system state and input information may induce safety risk \citep{PetitTITS2015,Mcdaniel2009}. For example, for cooperative control of multiple connected vehicles, the system state and input usually include vehicles' location and velocity messages, which should be well protected to prevent  adversaries from using such information to secretly track a vehicle \citep{Privacy_V2V,tracking} and from engaging in further malicious activities \citep{vehicle_privacy1,vehicle_privacy}. In its Readiness Report, the National Highway Traffic Safety Administration acknowledges various privacy issues that must be addressed when implementing vehicle communications, including preventing location tracking \citep{NHTSA_report}.
It is clear that the attacker can successfully eavesdrop the messages $x(k)$ and $u(k)$ when the conventional cloud-based MPC architecture introduced in Section \ref{subsec_cloud MPC} is adopted. The objective of this paper is to develop a masking mechanism to redesign the exchanged information between the plant and the cloud such that an equivalent MPC problem is solved without affecting system performance while preventing the external eavesdropper or untrusted cloud from inferring the system state and input.

%\vspace{-3mm}
\section{Main Results} \label{sec_main}
In this section, we present our privacy-preserving cloud-based nonlinear MPC framework. We first show that by applying affine masking on the states and controls, and transforming the cost terms and system dynamics accordingly, the transformed nonlinear MPC problem solved on the cloud is equivalent to the original problem. We then show that this affine transformation can protect the privacy of the system states and inputs by virtue of indistinguishability.

%\vspace{-3mm}
\subsection{Affine masking and problem reformulation} \label{subsection:affineMasking}
Inspired by the works \cite{Xu2015ACM,Xu2017,SultangazinTAC2021} that exploit linear transformations for \textit{linear} MPCs, in this section, we design affine transformation maps to accomplish the privacy protection for the considered cloud-based \textit{nonlinear} MPC. Specifically, two invertible affine maps $\iota_{x}(\cdot):=\left\lbrace P_{x}, t_{x} \right\rbrace$ and $\iota_{u}(\cdot):=\left\lbrace P_{u}, t_{u} \right\rbrace$ are introduced to transform the state $x(k)$ and input $u(k)$ to the new state $\bar{x}(k)$ and input $\bar{u}(k)$, as follows:
\begin{equation} \label{equ_affine}
	\begin{aligned}
		\bar{x}(k) &= \iota_{x}(x(k)) =  P_{x}(x(k)+t_{x}),
		\\
		\bar{u}(k) &= \iota_{u}(u(k)) =  P_{u}(u(k)+t_{u}),
	\end{aligned}
\end{equation}
where $P_{x}\in \mathbb{R}^{n\times n}$, $P_{u}\in \mathbb{R}^{m\times m}$ are arbitrary invertible matrices, and $t_{x}\in \mathbb{R}^{n}$, $t_{u}\in \mathbb{R}^{m}$ are arbitrary non-zero vectors with compatible dimensions. From \eqref{equ_system} and \eqref{equ_affine}, it follows that the transformed system state evolves according to the following dynamics:
\begin{equation} \label{equ_system_aff}
	\bar{x}(k+1) = \bar{\Phi}(\bar{x}(k), \bar{u}(k)) = \bar{f}(\bar{x}(k)) + \bar{g}(\bar{x}(k))\bar{u}(k),
\end{equation} 
where $\bar{f}(\cdot) \in \mathbb{R}^{n}$ and $\bar{g}(\cdot) \in \mathbb{R}^{n\times m}$ are defined as
\begin{equation} \label{equ_bar_f_bar_g}
	\begin{aligned}
		\bar{f}(\bar{x}(k)) &= P_{x}\left( f\circ \iota_{x}^{-1}(\bar{x}(k)) - g\circ \iota_{x}^{-1}(\bar{x}(k))t_{u} + t_{x} \right),
		\\
		\bar{g}(\bar{x}(k)) &= P_{x}g\circ \iota_{x}^{-1}(\bar{x}(k))P_{u}^{-1}, 
	\end{aligned}
\end{equation}
with $\circ$ denoting function composition and $\iota_{x}^{-1}(\cdot)$ being the inverse operation of $\iota_{x}(\cdot)$, i.e., $\iota_{x}^{-1}(\bar{x}(k))=P_{x}^{-1}\bar{x}(k)-t_{x}$. As will be shown below, the affine maps are able to mask the real system state $x(k)$ and input $u(k)$ to protect the privacy, and in the cloud a new optimization problem with respect to $\bar{x}(k)$, $\bar{u}(k)$, and the new system dynamics \eqref{equ_system_aff} are solved. Specifically, with the affine maps $\left\lbrace P_{x}, t_{x} \right\rbrace$ and $\left\lbrace P_{u}, t_{u} \right\rbrace$, one can show that $\textbf{P-1}$ can be transformed into the following problem:
\begin{equation} \label{equ_MPC_aff}
	%{\small
		\begin{aligned}
			\textbf{P-2}:\;\; &\min_{\bar{\bm{U}}_{k}} \bar{J}_{N}(\bar{x}(k), \bar{\bm{U}}_{k}) 
			\\
			&= \sum_{i=0}^{N-1}(\bar{x}_{i|k}^{\top}\bar{Q}\bar{x}_{i|k} + \bar{q}^{\top}\bar{x}_{i|k} + \bar{u}_{i|k}^{\top}\bar{R}\bar{u}_{i|k} + \bar{r}^{\top}\bar{u}_{i|k}) 
			\\
			&\;\;\;\; + \bar{x}_{N|k}^{\top}\bar{Q}_{f}\bar{x}_{N|k} + \bar{q}_{f}^{\top}\bar{x}_{N|k},
			\\
			\text{s.t.}\;\; & \bar{x}_{i+1|k} = \bar{f}(\bar{x}_{i|k}) + \bar{g}(\bar{x}_{i|k}) \bar{u}_{i|k}, i=0, \cdots, N-1,
			\\
			& \bar{x}_{i|k} \in \bar{\mathcal{X}}, i=1, \cdots, N-1,
			\\
			& \bar{u}_{i|k} \in \bar{\mathcal{U}}, i=0, \cdots, N-1,
			\\
			& \bar{x}_{N|k} \in \bar{\mathcal{X}}_{f},
			\\
			& \bar{x}_{0|k} = \bar{x}(k), \bar{\bm{U}}_{k}=\begin{bmatrix}
				\bar{u}_{0|k}^{\top}, \cdots, \bar{u}_{N-1|k}^{\top}
			\end{bmatrix}^{\top},
		\end{aligned}
		%}
\end{equation}
where $\bar{Q} \in \mathbb{R}^{n\times n}$, $\bar{q}\in \mathbb{R}^{n}$, $\bar{R}\in \mathbb{R}^{m\times m}$, $\bar{r}\in \mathbb{R}^{m}$, $\bar{Q}_{f}\in \mathbb{R}^{n\times n}$, and $\bar{q}_{f}\in \mathbb{R}^{n}$ are defined as
\begin{equation} \label{equ_para_aff}
	\begin{aligned}
		\bar{Q} &= P_{x}^{-\top}QP_{x}^{-1},\;\;\; \bar{q} = P_{x}^{-\top}q -2P_{x}^{-\top}Qt_{x},
		\\
		\bar{R} &= P_{u}^{-\top}RP_{u}^{-1},\;\;\; \bar{r} = P_{u}^{-\top}r -2P_{u}^{-\top}Rt_{u},
		\\
		\bar{Q}_{f} &= P_{x}^{-\top}Q_{f}P_{x}^{-1},\;\;\; \bar{q}_{f} = P_{x}^{-\top}q_{f} -2P_{x}^{-\top}Q_{f}t_{x}.
	\end{aligned}
\end{equation}
Moreover, in \eqref{equ_MPC_aff}, $\bar{\mathcal{X}}$, $\bar{\mathcal{X}}_{f}$ and $\bar{\mathcal{U}}$ are the corresponding constraint sets of $\mathcal{X}$, $\mathcal{X}_{f}$ and $\mathcal{U}$ under the affine maps $\left\lbrace P_{x}, t_{x} \right\rbrace$ and $\left\lbrace P_{u}, t_{u} \right\rbrace$, respectively. This indicates that $\forall x\in \mathcal{X}$, $\iota_{x}(x)=P_{x}(x+t_{x}) \in \bar{\mathcal{X}}$; vice versa $\forall \bar{x}\in \bar{\mathcal{X}}$, $\iota_{x}^{-1}(\bar{x})=P_{x}^{-1}\bar{x}-t_{x} \in \mathcal{X}$ (similarly for $\mathcal{X}_{f}$, $\bar{\mathcal{X}}_{f}$ and $\mathcal{U}$, $\bar{\mathcal{U}}$).

\begin{figure}[!t]
	\centering
	\hspace{0 in}
	\includegraphics[width=3.2 in]{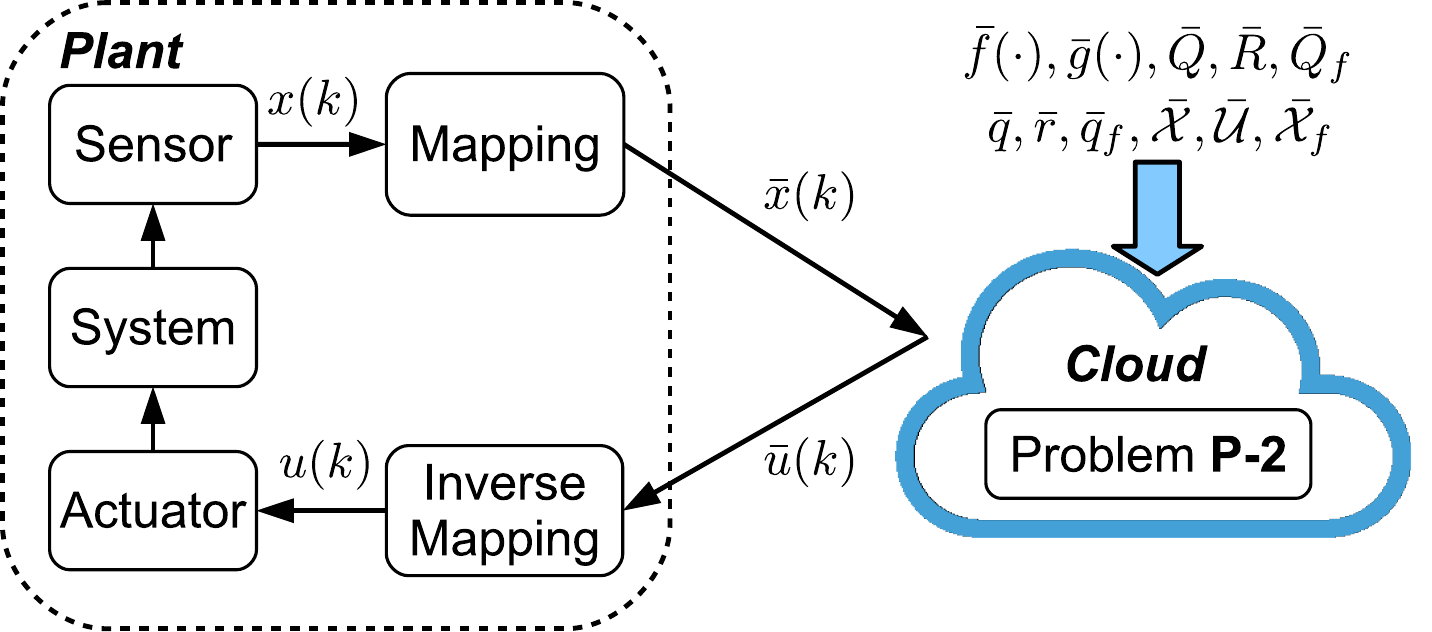}
	\vspace{1.5mm}
	\caption{Cloud-based privacy-preserving MPC architecture with affine masking.}
	\label{fig_secure MPC}
	%\vspace{-5mm}
\end{figure}

After introducing the affine maps, compared to the conventional cloud-based MPC in Section~II-B, our privacy-preserving cloud-based nonlinear MPC architecture is modified as shown in Fig.~\ref{fig_secure MPC}:
\begin{itemize}
	\item Handshaking Phase: Given the affine maps $\left\lbrace P_{x}, t_{x} \right\rbrace$ and $\left\lbrace P_{u}, t_{u} \right\rbrace$, the plant transforms its system dynamics, objective function and constraint sets into 
	\[
	\left\lbrace \bar{f}(\cdot), \bar{g}(\cdot), \bar{Q}, \bar{q}, \bar{R}, \bar{r}, \bar{Q}_{f}, \bar{q}_{f}, \bar{\mathcal{X}}, \bar{\mathcal{U}}, \bar{\mathcal{X}}_{f} \right\rbrace
	\] 
	and sends them to the cloud  to provide  necessary information for the cloud to set up the nonlinear programming problem \eqref{equ_MPC_aff}.
	
	\item Execution Phase: At each time step $k$, the plant first encodes $x(k)$ into $\bar{x}(k)$ with $\left\lbrace P_{x}, t_{x} \right\rbrace$ and sends $\bar{x}(k)$ to the cloud. Then the cloud computes $\bar{u}(k)$ by solving the optimization problem \eqref{equ_MPC_aff} and sends the solution $\bar{u}(k)$ to the plant. Finally, the plant uses $\left\lbrace P_{u}, t_{u} \right\rbrace$ to decode $\bar{u}(k)$, i.e., $u(k) = \iota_{u}^{-1}(\bar{u}(k)) = P_{u}^{-1}\bar{u}(k)-t_{u}$, and then applies the resultant $u(k)$ to the actuators. The system then evolves over one step.
\end{itemize}

\begin{remark}
	Compared to the conventional MPC architecture, the privacy-preserving MPC architecture requires the plant to mask the real state $x(k)$ into $\bar{x}(k)$ and decode $\bar{u}(k)$ into $u(k)$ by using the affine transformation. The affine transformation relies on matrix multiplication, whose time complexity is no greater than $O(n^3)$, where $n$ is the dimension of the transformed variables.
\end{remark}

Note that under the privacy-preserving cloud-based MPC architecture, the exchanged information between the plant and the cloud during the execution phase is $\bar{x}(k)$ and $\bar{u}(k)$, instead of the actual system state  $x(k)$ and input $u(k)$. In the sequel,  we first show that the transformed MPC problem solved on the cloud is equivalent to the original MPC problem, and we then show that the privacy of $x(k)$ and $u(k)$ is protected.  

\begin{assumption} \label{assumption1}
	Both the external eavesdropper and untrusted cloud can get access to the exchanged information between the plant and the cloud, i.e., $\bar{x}(k)$ and $\bar{u}(k)$, but they do not have any prior knowledge about the dynamic system, affine transformation scheme, and the affine maps $\left\lbrace P_{x}, t_{x} \right\rbrace$ and $\left\lbrace P_{u}, t_{u} \right\rbrace$.
\end{assumption}

\begin{lemma} \label{Lemma:affine}
	Under the affine transformation mechanism, the optimization problem $\textbf{P-2}$ is equivalent to $\textbf{P-1}$, i.e., if {\small$\bar{\bm{U}}^{*}_{k}=\begin{bmatrix}
			\bar{u}_{0|k}^{*\top}, \cdots, \bar{u}_{N-1|k}^{*\top} \end{bmatrix}^{\top}$} is a local (resp. global) minimizer of $\textbf{P-2}$, then the transformed control via inverse mapping {\small$\bm{U}^{*}_{k} =\begin{bmatrix} u_{0|k}^{*\top}, \cdots, u_{N-1|k}^{*\top} \end{bmatrix}^{\top} = \begin{bmatrix}
			(P_{u}^{-1}\bar{u}_{0|k}^{*}-t_{u})^{\top}, \cdots, (P_{u}^{-1}\bar{u}_{N-1|k}^{*}-t_{u})^{\top} 
		\end{bmatrix}^{\top}$} is a local (resp. global) minimizer of $\textbf{P-1}$. 
\end{lemma}

\begin{proof}
	Let {\small$\bar{\bm{X}}^{*}_{k}=\begin{bmatrix}
			\bar{x}_{0|k}^{*\top}, \cdots, \bar{x}_{N|k}^{*\top} \end{bmatrix}^{\top}$} and {\small$\bm{X}^{*}_{k}=\begin{bmatrix}
			x_{0|k}^{*\top}, \cdots, x_{N|k}^{*\top} \end{bmatrix}^{\top}$} be the state sequences corresponding to $\bar{\bm{U}}_{k}^{*}$ and $\bm{U}_{k}^{*}$, respectively. As $\bar{\bm{U}}_{k}^{*}$ is a minimizer of $\textbf{P-2}$, $\bar{\bm{X}}_{k}^{*}$ and $\bar{\bm{U}}_{k}^{*}$ satisfy the dynamic model \eqref{equ_system_aff} and the constraints described by $\left\lbrace \bar{\mathcal{X}}, \bar{\mathcal{U}}, \bar{\mathcal{X}}_{f} \right\rbrace$. 
	According to \eqref{equ_affine} and the formulation of problem $\textbf{P-1}$ and $\textbf{P-2}$, it can be concluded that if $\bm{U}_{k}^{*}$ is the inverse mapping of $\bar{\bm{U}}^{*}_{k}$ under $\left\lbrace P_{u}, t_{u} \right\rbrace$, then $\bm{X}_{k}^{*}$ and $\bm{U}_{k}^{*}$ are the state and input sequences of dynamic system \eqref{equ_system} and $\bm{X}_{k}^{*}$ is the inverse mapping of $\bar{\bm{X}}^{*}_{k}$ under $\left\lbrace P_{x}, t_{x} \right\rbrace$. In problem $\textbf{P-2}$, $\bar{\mathcal{X}}$, $\bar{\mathcal{X}}_{f}$ and $\bar{\mathcal{U}}$ are defined as the corresponding constraint sets of $\mathcal{X}$, $\mathcal{X}_{f}$ and $\mathcal{U}$ under the affine maps $\left\lbrace P_{x}, t_{x} \right\rbrace$ and $\left\lbrace P_{u}, t_{u} \right\rbrace$, respectively. Therefore, if $\bar{\bm{X}}_{k}^{*}$ and $\bar{\bm{U}}_{k}^{*}$ satisfy the constraints described by $\left\lbrace \bar{\mathcal{X}}, \bar{\mathcal{U}}, \bar{\mathcal{X}}_{f} \right\rbrace$, then $\bm{X}_{k}^{*}$ and $\bm{U}_{k}^{*}$ will satisfy the constraints described by $\left\lbrace \mathcal{X}, \mathcal{U}, \mathcal{X}_{f} \right\rbrace$. 
	% The above analysis indicates that 
	
	With our designed state and control transformations in \eqref{equ_affine},  the cost term transformations in \eqref{equ_para_aff}, and the definitions of $J_{N}(\cdot, \cdot)$ and $\bar{J}_{N}(\cdot, \cdot)$, it can be shown that 
	\begin{equation} \label{equ_J}
		J_{N}(x(k), \bm{U}_{k}) = \bar{J}_{N}(\bar{x}(k), \bar{\bm{U}}_{k}) + \varrho,
	\end{equation}
	where $\varrho = \sum_{i=1}^{N-1}\left( t_{x}^{\top}Qt_{x} -q^{\top}t_{x} + t_{u}^{\top}Rt_{u} - r^{\top}t_{u} \right) + t_{x}^{\top}Q_{f}t_{x}- q_{f}^{\top}t_{x} \in \mathbb{R}$ is a constant. We now use proof by contradiction, that is, we assume that $\bar{\bm{U}}_{k}^{*}$ is a local (resp. global) minimizer of problem $\textbf{P-2}$ within domain $\bar{\mathcal{U}}_{\text{local}}$ but $\bm{U}_{k}^{*}$ is not a local (resp. global) minimizer of problem $\textbf{P-1}$ within domain $\mathcal{U}_{\text{local}}$, where $\bar{\mathcal{U}}_{\text{local}}$ is the corresponding domain of $\mathcal{U}_{\text{local}}$ under the affine map $\left\lbrace P_{u}, t_{u} \right\rbrace$. This means that there exists an optimal sequence (other than $\bm{U}_{k}^{*}$) {\small$\bm{U}_{k}^{**} = \begin{bmatrix}
			u_{0|k}^{**\top}, \cdots, u_{N-1|k}^{**\top} \end{bmatrix}^{\top} \in \mathcal{U}_{\text{local}}$} such that
	\begin{equation} \label{equ_J_ineq}
		J_{N}(x(k), \bm{U}_{k}^{**}) < J_{N}(x(k), \bm{U}_{k}^{*}).
	\end{equation}
	Let {\small$\bar{\bm{U}}_{k}^{**} = \begin{bmatrix}
			\bar{u}_{0|k}^{**\top}, \cdots, \bar{u}_{N-1|k}^{**\top} \end{bmatrix}^{\top} = \begin{bmatrix}
			(P_{u}(u_{0|k}^{**}+t_{u}))^{\top}, \cdots, (P_{u}(u_{N-1|k}^{**}+t_{u}))^{\top} \end{bmatrix}^{\top}\in \bar{\mathcal{U}}_{\text{local}}$}. According to \eqref{equ_J}, \eqref{equ_J_ineq} can be rewritten as 
	\begin{equation}
		\bar{J}_{N}(\bar{x}(k), \bar{\bm{U}}_{k}^{**}) + \varrho < \bar{J}_{N}(\bar{x}(k), \bar{\bm{U}}_{k}^{*}) + \varrho,
	\end{equation}
	which contradicts the assumption that $\bar{\bm{U}}_{k}^{*}$ is a local (resp. global) minimizer of problem $\textbf{P-2}$. The proof is complete.	
\end{proof}

Lemma~\ref{Lemma:affine} reveals that the transformed MPC problem is a different yet equivalent form of the original MPC problem. Thus, if the original MPC ensures properties such as recursive feasibility and asymptotical stability, then the transformed formulation preserves these theoretical guarantees.

%\vspace{-3mm}
\subsection{Privacy Preservation}
We next discuss the privacy notion used in this paper. As mentioned in the previous section, the attacker aims to infer the system state $x(k)$ and control input $u(k)$. Under the privacy-preserving cloud-based MPC architecture discussed above, the attacker will have access to $\bar{x}(k)$ and $\bar{u}(k)$ at each time step $k$, and we need to show that for any $\kappa\in \mathbb{N}_{+}$, $x_{[0,\kappa]} = \lbrace x(0), \cdots, x(\kappa) \rbrace$ and $u_{[0,\kappa]} = \lbrace u(0), \cdots, u(\kappa) \rbrace$ cannot be identified from $\bar{x}_{[0,\kappa]} = \lbrace \bar{x}(0), \cdots, \bar{x}(\kappa) \rbrace$ and $\bar{u}_{[0,\kappa]} = \lbrace \bar{u}(0), \cdots, \bar{u}(\kappa) \rbrace$. To facilitate the following development, two triples $\Omega$ and $\bar{\Omega}$ are defined as
\begin{equation}
	\begin{aligned}
		\Omega &= \left\lbrace \left\lbrace f(\cdot), g(\cdot) \right\rbrace, J_{N}(\cdot, \cdot),  \left\lbrace \mathcal{X}, \mathcal{U}, \mathcal{X}_{f} \right\rbrace\right\rbrace,
		\\
		\bar{\Omega} &= \left\lbrace \left\lbrace \bar{f}(\cdot), \bar{g}(\cdot) \right\rbrace, \bar{J}_{N}(\cdot, \cdot),  \left\lbrace \bar{\mathcal{X}}, \bar{\mathcal{U}}, \bar{\mathcal{X}}_{f} \right\rbrace\right\rbrace.
	\end{aligned}
\end{equation}
It can be seen that the triples $\Omega$ and $\bar{\Omega}$ can be used to define the optimization problem in \eqref{equ_MPC} and \eqref{equ_MPC_aff}, respectively. 
%We call $\left\lbrace x_{[0,\kappa]}, u_{[0,\kappa]} \right\rbrace$ a solution to the optimization problem \eqref{equ_MPC} defined by $\Omega$ if $\left\lbrace x_{[0,\kappa]}, u_{[0,\kappa]} \right\rbrace$ is a trajectory of the nonlinear system $\left\lbrace f(\cdot), g(\cdot) \right\rbrace$ minimizing objective function $J_{N}(\cdot, \cdot)$ under constraints described by $\left\lbrace \mathcal{X}, \mathcal{U}, \mathcal{X}_{f} \right\rbrace$. 
We call $\left\lbrace x_{[0,\kappa]}, u_{[0,\kappa]} \right\rbrace$ a solution to the optimization problem \eqref{equ_MPC} defined by $\Omega$ if $\left\lbrace x_{[0,\kappa]}, u_{[0,\kappa]} \right\rbrace$ is a trajectory of the nonlinear system $\left\lbrace f(\cdot), g(\cdot) \right\rbrace$ where the control input $u(k)$ at each time step $k$ is solved by minimizing objective function $J_{N}(\cdot, \cdot)$ under constraints described by $\left\lbrace \mathcal{X}, \mathcal{U}, \mathcal{X}_{f} \right\rbrace$.
Moreover, we use {\small$\Omega \xRightarrow{\lbrace P_{x}, t_{x}, P_{u}, t_{u} \rbrace} \bar{\Omega}$} to denote that $\bar{\Omega}$ is the transformed triple of $\Omega$ under the affine maps $\left\lbrace P_{x}, t_{x} \right\rbrace$ and $\left\lbrace P_{u}, t_{u} \right\rbrace$.

Given $\bar{\Omega}$, for any feasible input sequence $\bar{x}_{[0,\kappa]}$ and output sequence $\bar{u}_{[0,\kappa]}$ received by the attacker, the set $\Delta_{\bar{\Omega}}(\bar{x}_{[0,\kappa]}, \bar{u}_{[0,\kappa]})$ is defined as
\begin{equation}
	\begin{aligned}
		&\Delta_{\bar{\Omega}}(\bar{x}_{[0,\kappa]}, \bar{u}_{[0,\kappa]}) = \lbrace x_{[0,\kappa]}, u_{[0,\kappa]}: \exists \lbrace P_{x}, t_{x}, P_{u}, t_{u} \rbrace \; \text{and}\; \Omega
		\\
		&\text{s.t.} \; \bar{x}(k)=P_{x}(x(k)+t_{x}), \bar{u}(k)=P_{u}(u(k)+t_{u}), 
		%\forall k=0, \cdots, \kappa
		\\
		&\Omega \xRightarrow{\lbrace P_{x}, t_{x}, P_{u}, t_{u} \rbrace} \bar{\Omega}, \text{and} \left\lbrace x_{[0,\kappa]}, u_{[0,\kappa]} \right\rbrace \text{is the solution to}\; \Omega \rbrace.
	\end{aligned}
\end{equation}
Essentially, the set $\Delta_{\bar{\Omega}}(\bar{x}_{[0,\kappa]}, \bar{u}_{[0,\kappa]})$ includes all possible values of $\lbrace x_{[0,\kappa]}, u_{[0,\kappa]} \rbrace$ that can be transformed into $\lbrace \bar{x}_{[0,\kappa]}, \bar{u}_{[0,\kappa]} \rbrace$ with corresponding affine maps $\lbrace P_{x}, t_{x}, P_{u}, t_{u} \rbrace$. The diameter of $\Delta_{\bar{\Omega}}(\bar{x}_{[0,\kappa]}, \bar{u}_{[0,\kappa]})$, a metric that measures the distance (dissimilarity) between its elements, is defined as
\begin{equation} \label{equ_diam}
	\text{Diam}_{\Delta_{\bar{\Omega}}}(\bar{x}_{[0,\kappa]}, \bar{u}_{[0,\kappa]}) = \sup_{w, w'\in \Delta_{\bar{\Omega}}(\bar{x}_{[0,\kappa]}, \bar{u}_{[0,\kappa]})} \left| w-w' \right|_{\min},
\end{equation}
where $\left| w-w' \right|_{\min} = \min_{l\in \left\lbrace 1, \cdots, (n+m)(\kappa+1) \right\rbrace} \left| w_{l}-w'_{l} \right|$ with $w_{l}$ and $w'_{l}$ being the $l$-th element of $w$ and $w'$, respectively. Note that $\left| w-w' \right|_{\min}$ is used to quantify the minimum element difference between $w$ and $w'$. If $\left| w-w' \right|_{\min} = \delta$, where $\delta$ is an arbitrarily positive constant, then $\forall l\in \left\lbrace 1, \cdots, (n+m)(\kappa+1) \right\rbrace$, we have $\left| w_{l}-w'_{l} \right|\ge \delta$.

\begin{definition} [$\infty$-Diversity with Unbounded Diameter] \label{def_privacy}
	The privacy of the actual system state $x_{[0,\kappa]}$ and input $u_{[0,\kappa]}$ is preserved if $\left. 1 \right)$ the cardinality of the set $\Delta_{\bar{\Omega}}(\bar{x}_{[0,\kappa]}, \bar{u}_{[0,\kappa]})$ is infinite, and $\left. 2 \right)$ $\text{Diam}_{\Delta_{\bar{\Omega}}}(\bar{x}_{[0,\kappa]}, \bar{u}_{[0,\kappa]}) = \infty$.  
\end{definition}

In the $\infty$-Diversity with Unbounded Diameter privacy defined above, the first condition requires that there are infinitely many sets of $\lbrace x_{[0,\kappa]}, u_{[0,\kappa]} \rbrace$, $\lbrace P_{x}, t_{x}, P_{u}, t_{u} \rbrace$ and $\Omega$ that can generate the same $\lbrace \bar{x}_{[0,\kappa]}, \bar{u}_{[0,\kappa]} \rbrace$ received by the attacker. As a result, it is impossible for the attacker to use $\lbrace \bar{x}_{[0,\kappa]}, \bar{u}_{[0,\kappa]} \rbrace$ to infer the actual system state and input information. Moreover, the second condition requires that the difference between the possible values of each element in $\lbrace x_{[0,\kappa]}, u_{[0,\kappa]} \rbrace$ could be arbitrarily large, and thus the attacker cannot even approximately estimate (e.g., find a finite range or uniquely determine a portion of) the private signals.

We now show that the affine transformation mechanism can achieve privacy preservation based on Definition~\ref{def_privacy}.
\begin{theorem} \label{theorem_privacy}
	Under the affine masking mechanism described in Section~\ref{subsection:affineMasking}, the system states and control inputs are $\infty$-diversity-with-unbounded-diameter private, that is, the attacker cannot infer the actual system state $x(k)$ and input $u(k)$ with any guaranteed accuracy.
\end{theorem}

\begin{proof}
	We prove Theorem \ref{theorem_privacy} by proving the two conditions in Definition 1. We first show that under the affine masking scheme, the cardinality of the set $\Delta(\bar{x}_{[0,\kappa]}, \bar{u}_{[0,\kappa]})$ is infinite. Specifically, given the sequence $\lbrace \bar{x}_{[0,\kappa]}, \bar{u}_{[0,\kappa]} \rbrace$ and $\bar{\Omega}$ accessible to the attacker, for arbitrary affine maps $\iota_{x}'(\cdot):=\lbrace P_{x}', t_{x}'\rbrace$ and $\iota_{u}'(\cdot):=\lbrace P_{u}', t_{u}' \rbrace$ such that $ P_{x}'$ and $P_{u}'$ are invertible, a sequence $\lbrace x'_{[0,\kappa]}, u'_{[0,\kappa]} \rbrace$ and $\Omega'$ can be uniquely determined. %based on $\lbrace \bar{x}_{[0,\kappa]}, \bar{u}_{[0,\kappa]} \rbrace$ and $\bar{\Omega}$ by using $\lbrace P_{x}', t_{x}', P_{u}', t_{u}' \rbrace$ as an inverse mapping. 
	Recall that $\lbrace x'_{[0,\kappa]}, u'_{[0,\kappa]} \rbrace$ should satisfy $\bar{x}(k)=P_{u}'(x'(k)+t_{x}')$ and $\bar{u}(k)=P_{u}'u_{k}'+t_{u}'$, which indicates that the sequence $\lbrace x'_{[0,\kappa]}, u'_{[0,\kappa]} \rbrace$ can be determined by 
	\begin{equation} \label{equ_xk_prime}
		\begin{aligned}
			x'(k) = \iota_{x}'^{-1}(\bar{x}(k)) = (P_{x}')^{-1}\bar{x}(k)-t'_{x},
			\\
			u'(k) = \iota_{u}'^{-1}(\bar{u}(k)) = (P_{u}')^{-1}\bar{u}(k)-t'_{u}.
		\end{aligned}
	\end{equation} 
	Based on \eqref{equ_xk_prime} and $\bar{\Omega}$, $\Omega'$ can be further obtained by following the similar procedure introduced in Section~\ref{subsection:affineMasking}.
	As there exist infinitely many such affine maps $\lbrace P_{x}', t_{x}', P_{u}', t_{u}' \rbrace$, there exist infinitely many $\lbrace x'_{[0,\kappa]}, u'_{[0,\kappa]} \rbrace$ and $\Omega'$ such that via proper affine transformations, the attacker will receive the same accessed information: $\lbrace \bar{x}_{[0,\kappa]}, \bar{u}_{[0,\kappa]} \rbrace$ and $\bar{\Omega}$, which thus satisfies the first condition in Definition~\ref{def_privacy}.

	We now prove the second condition in Definition~\ref{def_privacy}. For any $w, w' \in \Delta_{\bar{\Omega}}(\bar{x}_{[0,\kappa]}, \bar{u}_{[0,\kappa]})$ (i.e., $\lbrace x_{[0,\kappa]}, u_{[0,\kappa]} \rbrace, \lbrace x'_{[0,\kappa]}, u'_{[0,\kappa]} \rbrace \in \Delta_{\bar{\Omega}}(\bar{x}_{[0,\kappa]}, \bar{u}_{[0,\kappa]})$) with $\lbrace P_{x}, t_{x}, P_{u}, t_{u} \rbrace$ and $\lbrace P_{x}', t_{x}', P_{u}', t_{u}' \rbrace$ being the corresponding affine maps, we have 
	%$\bar{x}(k) = P_{x}(x(k)+t_{x}) = P_{x}'(x'(k)+t'_{x})$ and $\bar{u}(k) = P_{u}(u(k)+t_{u}) = P_{u}'(u'(k)+t'_{u})$, indicating that
	\begin{equation} \label{equ_xk_uk}
		\begin{aligned}
			x(k) = P_{x}^{-1}\bar{x}(k)-t_{x}, \;\; x'(k) = (P_{x}')^{-1}\bar{x}(k)-t'_{x},
			\\
			u(k) = P_{u}^{-1}\bar{u}(k)-t_{u}, \;\; u'(k) = (P_{u}')^{-1}\bar{u}(k)-t'_{u}.
		\end{aligned}		
	\end{equation}
	Based on \eqref{equ_xk_uk}, it can be obtained that
	\begin{equation} \label{equ_w}
		%\small{
			\begin{aligned}
				&\left| w-w' \right|_{\min} = \left| \begin{matrix}
					(P_{x}^{-1}-(P_{x}')^{-1})\bar{x}(0)-(t_{x}-t_{x}')
					\\
					\vdots
					\\
					(P_{x}^{-1}-(P_{x}')^{-1})\bar{x}(\kappa)-(t_{x}-t_{x}')
					\\
					(P_{u}^{-1}-(P_{u}')^{-1})\bar{u}(0)-(t_{u}-t_{u}')
					\\
					\vdots
					\\
					(P_{u}^{-1}-(P_{u}')^{-1})\bar{u}(\kappa)-(t_{u}-t_{u}')
				\end{matrix} \right|_{\min}
				\\
				&= \left| \begin{matrix} 
					(I_{\kappa+1}\otimes (P_{x}^{-1}-(P_{x}')^{-1}))\bar{x}_{[0, \kappa]}-1_{\kappa+1}\otimes (t_{x}-t_{x}')
					\\
					(I_{\kappa+1}\otimes (P_{u}^{-1}-(P_{u}')^{-1}))\bar{u}_{[0,\kappa]}-1_{\kappa+1}\otimes (t_{u}-t_{u}')
				\end{matrix} \right|_{\min}
				\\
				&\ge \left| \begin{matrix}
					1_{\kappa+1}\otimes (t_{x}-t_{x}')
					\\
					1_{\kappa+1}\otimes (t_{u}-t_{u}')
				\end{matrix} \right|_{\min} - \left| \begin{matrix} 
					(I_{\kappa+1}\otimes (P_{x}^{-1}-(P_{x}')^{-1}))\bar{x}_{[0, \kappa]}
					\\
					(I_{\kappa+1}\otimes (P_{u}^{-1}-(P_{u}')^{-1}))\bar{u}_{[0,\kappa]}
				\end{matrix} \right|_{\max},
			\end{aligned}
			%}
	\end{equation}
	where $\otimes$ is the Kronecker product, $I_{\kappa+1}\in \mathbb{R}^{(\kappa+1)\times (\kappa+1)}$ is the identity matrix, and $1_{\kappa+1} \in \mathbb{R}^{\kappa+1}$ is the column vector with all the entries being ones. Furthermore, by using \eqref{equ_diam} and \eqref{equ_w}, the diameter of the set $\Delta_{\bar{\Omega}}(\bar{x}_{[0,\kappa]}, \bar{u}_{[0,\kappa]})$ can be derived as follows:
	\begin{equation}		
		%\small{
			\begin{aligned}
				&\text{Diam}_{\Delta_{\bar{\Omega}}}(\bar{x}_{[0,\kappa]}, \bar{u}_{[0,\kappa]}) = \sup_{w, w'\in \Delta_{\bar{\Omega}}(\bar{x}_{[0,\kappa]}, \bar{u}_{[0,\kappa]})} \left| w-w' \right|_{\min}
				\\
				&\ge \sup_{t_{x}, t_{x}'\in \mathbb{R}^{n}, t_{u}, t_{u}' \in \mathbb{R}^{m}} \left| \begin{matrix}
					1_{\kappa+1}\otimes (t_{x}-t_{x}')
					\\
					1_{\kappa+1}\otimes (t_{u}-t_{u}')
				\end{matrix} \right|_{\min}
				\\
				& \;\;\;\; -\inf_{P_{x}, P_{x}'\in \mathbb{R}^{n\times n}, P_{u}, P_{u}'\in \mathbb{R}^{m\times m}} \left| \begin{matrix} 
					(I_{\kappa+1}\otimes (P_{x}^{-1}-(P_{x}')^{-1}))\bar{x}_{[0, \kappa]}
					\\
					(I_{\kappa+1}\otimes (P_{u}^{-1}-(P_{u}')^{-1}))\bar{u}_{[0,\kappa]}
				\end{matrix} \right|_{\max}
				\\
				&= \infty.
			\end{aligned}
			%}
	\end{equation}
	Thus, the second condition in Definition~\ref{def_privacy} is satisfied. 
\end{proof}

By following the arguments from the proof of Theorem~1, it is clear that there exist infinitely many sets of $\Omega$ (i.e., system dynamics, cost function, and constraint sets) such that via proper affine transformation, the attacker will receive the same accessed information $\bar{\Omega}$. Therefore, the attacker cannot exploit $\bar{\Omega}$ to uniquely determine the actual $\Omega$. Due to complicated structure of $\Omega$, defining metrics to quantify the difference between the accessible valuations of $\Omega$ is non-trivial and needs to be further studied.

\begin{remark}
	Due to communication overhead or resource constraint, elements in the affine maps $\left\lbrace P_{x}, t_{x} \right\rbrace$ and $\left\lbrace P_{u}, t_{u} \right\rbrace$ cannot be arbitrary large numbers in practical applications. To disguise the real state and input information, it is beneficial for the plant to choose suitable affine maps such that the transformed data $\left\lbrace \bar{x}_{[0,\kappa]}, \bar{u}_{[0,\kappa]} \right\rbrace$ is quite different from the actual one $\left\lbrace x_{[0,\kappa]}, u_{[0,\kappa]} \right\rbrace$. Generally, within a bounded set confined by communication overhead or resource constraint, the plant can choose $P_{x}$, $P_{u}$ and $t_{x}$, $t_{u}$ that are distant (in the sense of Frobenius norm, for example) from the identity matrix and zero vector, respectively, to achieve this purpose.
\end{remark}

\subsection{Discussion on Privacy Notion and Protection Scheme}
Definition~\ref{def_privacy} is an extension to the $l$-diversity \citep{machanavajjhala2007} which has been widely adopted in formal privacy analysis on attribute privacy of tabular datasets and has recently been extended to define privacy in linear dynamic networks \citep{Lu2020}. Essentially, $l$-diversity requires that there are at least $l$ different possible values for the privacy sensitive data attributes, and a greater $l$ indicates greater indistinguishability. Definition 1 extends the $l$-diversity notion by requiring that there exist infinitely many possible sets of states/inputs and affine transformation combinations that can generate the same accessible information for the adversary ($\infty$-diversity). In addition, the difference of the states/inputs in these sets can be arbitrarily large (unbounded diameter). This makes the adversary unable to identify the actual value or even estimate a rough range or a portion of the private parameters. Furthermore, the conventional $l$-diversity works for discrete-valued setting, whereas Definition 1 is tailored to the considered cloud-based nonlinear MPC with sensitive attributes being continuous-valued. In the following, we discuss the differences between the proposed privacy definition/scheme and other existing  privacy  notions/schemes (e.g., differential privacy, homomorphic encryption, and affine transformation). 

The privacy notions based on statistics or information theory have been widely utilized in the security community, such as differential privacy, entropy, and mutual information. Differential privacy approaches inject random noises into private data in such a way that the adversary cannot infer the private data with high probability \citep{Dwork2014,huang2012ACM}. For cloud-based nonlinear MPC, such persistent noise injection mechanism will inevitably deteriorate system performance and potentially lead to the violation of state and input constraints, while the proposed privacy definition with the affine masking scheme does not affect system performance as the transformed problem is equivalent to the original one as shown in Section~\ref{subsection:affineMasking}. Moreover, both entropy and mutual information-based privacy preservation relies on explicit statistical models of source data and side information \citep{Nekouei2019ARC,Sankar2013}, which, however, are not generally available in the considered cloud-based MPC problem as the state and input signals of the system may not follow any probabilistic distribution.

%Compared to differential privacy that injects independent noises to obfuscate private values that will inevitably compromise algorithmic accuracy \citep{Dwork2014,huang2012ACM}, the considered privacy definition with the affine masking scheme does not affect system performance as the transformed problem is equivalent to the original one as shown in Section~\ref{subsection:affineMasking}. Entropy and mutual information-based privacy preservation relies on explicit statistical models of source data and side information \citep{Nekouei2019ARC,Sankar2013} which, however, are not generally available in the considered cloud-based MPC problem. 
%Semantic security requires ``nothing is learned'' by the adversary from its accessible data, which intrinsically inhibits any meaningful data utility and can resist arbitrary side information \cite{Dwork2014,Lu2020}. In our problem, the eavesdropper has access to all exchanged information that is necessary for the cloud to perform the optimizations. Therefore, semantic security is too restrictive and not applicable to our problem.

Various homomorphic encryption-based methods have been designed for privacy-preserving linear MPC, and both semantic security and secret sharing have been used to define privacy.
%Semantic security requires that no additional information about a plaintext can be inferred using its ciphertext by the adversary, and some encryption systems with semantic security guarantees have recently been designed for cloud-based MPC \citep{AlexandruCDC2018,DarupIFAC2018}. 
Semantic security requires that no additional information about a plaintext can be inferred using its ciphertext by the adversary. It is worth noting that the encryption techniques with semantic security guarantees \citep{NilsCDC2020,SchulzeCSL2018,AlexandruCDC2018,DarupIFAC2018} only allow the cloud (which has the public key but not the private key) to perform simple linear mathematical operations on encrypted data,
%. However, these approaches cannot be directly used for nonlinear systems considered in this manuscript.
making them applicable only for linear systems and difficult, if not impossible, to be extended to the considered nonlinear system with complicated operations.
In addition, secret sharing allows to divide and reconstruct secret data in such a way that the individual shareholders reveal nothing about the secret \citep{Shamir1979}. It is an effective tool to achieve privacy-preserving cloud-based control \citep{Darup2019CDC,Schlor2021CDC} but it requires using multiple shareholders/clouds that are not colluding, resulting in a more complex system structure. Instead of relying on data division and sharing to multiple shareholders, the proposed method exploits affine transformation to mask sensitive system information, which does not require multiple clouds to facilitate the design of privacy preservation scheme. 

The proposed affine masking scheme is inspired by the algebraic transformation-based works designed for linear systems %\citep{Xu2015ACM,Xu2017,SultangazinTAC2021,Naseri2022ECC} 
but there exist several differences. In \cite{Xu2015ACM}, the linear MPC problem with linear input constraints is first transformed into a quadratic problem, and then a transformation mechanism based on non-singular matrices is designed to mask sensitive information. \cite{Xu2017} combines orthogonal matrices with homomorphic encryption to design a hybrid privacy preservation scheme for non-constrained linear MPC. Note that their transformation mechanisms are designed for specific linear MPC forms and thus cannot be applied to the considered nonlinear MPC with state and input constraints. In \cite{SultangazinTAC2021}, the dimension of the manifold describing the diversity experienced by the adversary is used as a measure of privacy. The derivation of the set dimension based privacy notion relies on the system's linear characteristics, and thus it is difficult (if not impossible) to extend this notion to nonlinear systems. In this paper, we exploit the set cardinality and diameter to quantify the privacy for cloud-based nonlinear MPC, which applies to general nonlinear systems with constraints and can provide stronger privacy guarantees. Furthermore, in \cite{Naseri2022ECC}, the transformation-based technique is incorporated into set-theoretic MPC to protect the privacy of a linear system subject to bounded disturbance, while no rigorous notion is introduced to analyze the privacy guarantees. Our work focuses on privacy preservation in nonlinear MPC and the development of the privacy notion.

Although the proposed method circumvents some issues that arise in existing privacy notions, it has certain limitations. 
%One limitation is that it cannot resist arbitrary side information compared to differential privacy and semantic security.
One limitation is that it does not consider the case in which the external eavesdropper or untrusted cloud has auxiliary information about the dynamic system and the affine transformation scheme (see Assumption~\ref{assumption1}), whereas differential privacy and semantic security are immune to arbitrary auxiliary information.

\begin{remark}
	In summary, the proposed affine masking strategy for nonlinear state-feedback MPC makes two technical contributions. First, we tailor the affine masking technique to conceal sensitive information and reformulate the original nonlinear MPC into an equivalent formulation, achieving privacy preservation without compromising control performance. Different from homomorphic encryption-based methods that are limited to linear MPC and incur tedious encryption and decryption procedures, the proposed strategy is applicable to a class of control-affine nonlinear systems and is computationally efficient. Second, we introduce a new privacy definition that uses both set cardinality and diameter to facilitate the privacy quantification for nonlinear MPC. Existing transformation-based approaches rely on linear system characteristics, while our privacy notion extends the existing approaches to preserve privacy for nonlinear systems and employs the set cardinality and diameter to measure the uncertainties on each element of interest to the adversary, making it applicable to nonlinear systems with constraints.
	%as measures of uncertainty for each element of interest to potential adversaries.
\end{remark}

%\vspace{-3mm}
\section{Extension to Output-feedback MPC}	
The aforementioned cloud-based MPC methods require that all system states are measurable to perform the state-feedback control. However, for some systems, not all states are accessible but an output vector is available for output feedback control designs. Therefore, in this section we extend the privacy-preserving cloud-based MPC design to the output-feedback case. Specifically, let $y(k) \in \mathbb{R}^{p}$ be the system output described by
\begin{equation} \label{equ_y}
	y(k) = Cx(k),
\end{equation}
where $C\in \mathbb{R}^{p\times n}$. We assume that the system is observable and the state $x(k)$ can be estimated via a high-gain observer \citep{khalil2002nonlinear} in the following form:
\begin{equation} \label{equ_hat_x}
	\hat{x}(k+1) = \Phi(\hat{x}(k), u(k)) + H(y(k)-C\hat{x}(k)),
\end{equation}
where $\hat{x}(k)\in \mathbb{R}^{n}$ is the estimate of $x(k)$ and $H \in \mathbb{R}^{n\times p}$ is the gain matrix. The estimated state $\hat{x}(k)$ is then fed into the MPC problem (\ref{equ_MPC}) to obtain the solutions. Under the output-feedback case, the conventional cloud-based MPC is typically implemented as follows:
\begin{itemize}
	\item Handshaking Phase: The plant sends 
	\[
	\left\lbrace f(\cdot), g(\cdot), Q, q, R, r, Q_{f}, q_{f}, \mathcal{X}, \mathcal{U}, \mathcal{X}_{f}, H, C \right\rbrace
	\] 
	to the cloud, which are necessary information for the cloud to perform state estimation and subsequent MPC based on the estimated state.
	
	\item Execution Phase: At each time step $k$, the plant first sends $y(k)$ to the cloud. Then the cloud estimates the system state $\hat{x}_k$ via \eqref{equ_hat_x}, computes $u(k)$ based on $\hat{x}_k$ by solving the optimization problem shown in \eqref{equ_MPC} and sends $u(k)$ to the plant. Finally, the plant applies $u(k)$ to the actuators and the system evolves over one step.
\end{itemize} 

The objective now is to avoid leaking the privacy-sensitive information $y(k)$, $\hat{x}(k)$, and $u(k)$ to the attacker. Similar to \eqref{equ_affine}, an invertible affine map is introduced to mask $y(k)$ as follows:
\begin{equation} \label{equ_affine_y}
	\bar{y}(k) = \iota_{y}(y(k)) =  P_{y}(y(k)+t_{y}),
\end{equation}
where $P_{y}\in \mathbb{R}^{p\times p}$ is an invertible matrix and $t_{y}\in \mathbb{R}^{p}$ is an offset vector. According to \eqref{equ_affine}, \eqref{equ_y} and $\eqref{equ_affine_y}$, it can be obtained that 
\begin{equation} \label{equ_bar_y}
	\bar{y}(k) = \bar{C}\bar{x}(k) + \bar{\sigma}_{y},
\end{equation}
with $\bar{C} \in \mathbb{R}^{p\times n}$ and $\bar{\sigma}_{y} \in \mathbb{R}^{p}$ being defined as
\begin{equation} \label{equ_bar_C}
	\begin{aligned}
		\bar{C} &= P_{y}CP_{x}^{-1},
		\\
		\bar{\sigma}_{y} &= P_{y}(-Ct_{x}+t_{y}).
	\end{aligned}
\end{equation}
Moreover, from \eqref{equ_system_aff}, \eqref{equ_hat_x} and \eqref{equ_bar_y}, it can be shown that $\bar{x}(k)$ can be estimated with $\bar{y}(k)$ via the following observer:
\begin{equation} \label{equ_hat_bar_x}
	\hat{\bar{x}}(k+1) = \bar{\Phi}(\hat{\bar{x}}(k), \bar{u}(k)) + \bar{H}(\bar{y}(k)-\bar{C}\hat{\bar{x}}(k)-\bar{\sigma}_{y}),
\end{equation}
where $\bar{H} \in \mathbb{R}^{n\times p}$ is given by
\begin{equation}
	\bar{H} = P_{x}HP_{y}^{-1}.
\end{equation}
The cloud-based privacy-preserving MPC under the output-feedback setup can then be performed with the following modified procedures: 
\begin{itemize}
	\item Handshaking Phase: Given the affine maps $\left\lbrace P_{x}, t_{x} \right\rbrace$, $\left\lbrace P_{u}, t_{u} \right\rbrace$ and $\left\lbrace P_{y}, t_{y} \right\rbrace$, the plant transforms its system dynamics, objective function, constraint sets and observer into %$\left\lbrace \bar{f}(\cdot), \bar{g}(\cdot) \right\rbrace$, $\bar{J}_{N}(\cdot, \cdot)$ (i.e., $\left\lbrace \bar{Q}, \bar{q}, \bar{R}, \bar{r}, \bar{Q}_{f}, \bar{q}_{f} \right\rbrace$), $\left\lbrace \bar{\mathcal{X}}, \bar{\mathcal{U}}, \bar{\mathcal{X}}_{f} \right\rbrace$ and $\left\lbrace \bar{H}, \bar{C}, \bar{\sigma}_{y} \right\rbrace$ 
	\[
	\left\lbrace \bar{f}(\cdot), \bar{g}(\cdot), \bar{Q}, \bar{q}, \bar{R}, \bar{r}, \bar{Q}_{f}, \bar{q}_{f}, \bar{\mathcal{X}}, \bar{\mathcal{U}}, \bar{\mathcal{X}}_{f}, \bar{H}, \bar{C}, \bar{\sigma}_{y} \right\rbrace
	\] 
	and sends them to the cloud.
	
	\item Execution Phase: At each time step $k$, the plant first encodes $y(k)$ into $\bar{y}(k)=P_{y}(y(k)+t_{y})$ and sends $\bar{y}(k)$ to the cloud. Then the cloud estimates the system state via \eqref{equ_hat_bar_x}, computes $\bar{u}(k)$ by solving the optimization problem  \eqref{equ_MPC_aff} and sends $\bar{u}(k)$ to the plant. Finally, the plant uses $\left\lbrace P_{u}, t_{u} \right\rbrace$ to decode $\bar{u}(k)$ (i.e., $u(k) = \iota_{u}^{-1}(\bar{u}(k)) = P_{u}^{-1}\bar{u}(k)-t_{u}$) and then applies $u(k)$ to the actuators. The system evolves over one step.
\end{itemize}

\begin{theorem} \label{theorem_privacy_outupt}
	Under the affine masking mechanism described in this subsection, the system outputs, states and control inputs are $\infty$-diversity-with-unbounded-diameter private, that is,  the attacker cannot infer the actual outputs $y(k)$, system state $x(k)$ and input $u(k)$ with any guaranteed accuracy.
\end{theorem}

\begin{proof}
	The proof follows similar arguments in Theorem~\ref{theorem_privacy}. 
\end{proof}

\begin{remark}
	In contrast to Section~\ref{sec_main}, which employs affine maps to conceal real state and input information in state-feedback MPC, the privacy preservation scheme for output-feedback MPC introduces an additional affine map to mask the real system output. This process also entails reformulating the original high-gain observer into a compatible form, enabling estimation of the transformed system state with the transformed output. The combination of the affine masking strategy and observer reformulation is crucial to ensure that the original output-feedback MPC is shaped into a different but equivalent one, which guarantees that the private information is protected with no performance degradation.
\end{remark}

\section{Simulation Results}
In this section, we perform numerical simulations to demonstrate the efficacy of the developed approach. All computations are performed in MATLAB 2022a on a laptop with an Intel i7-10710U CPU with 6 cores, 1.6 GHz clock rate, and 16 GB RAM.
We consider the regulation control problem of a quadrotor aerial vehicle. The system state and input of the quadrotor aerial vehicle are defined as $x=\begin{bmatrix}
	\xi^{\top}, \eta^{\top}, \dot{\xi}^{\top}, \dot{\eta}^{\top}
\end{bmatrix}^{\top} \in \mathbb{R}^{12}$ and $u=\begin{bmatrix}
	u_{1}, u_{2}, u_{3}, u_{4}
\end{bmatrix}^{\top} \in \mathbb{R}^{4}$, respectively, where $\xi = \begin{bmatrix}
	\xi_{x}, \xi_{y}, \xi_{z}
\end{bmatrix}^{\top} \in \mathbb{R}^{3}$ represents the position of the quadrotor mass center expressed in the inertial frame, $\eta = \begin{bmatrix}
	\phi, \theta, \psi
\end{bmatrix}^{\top} \in \mathbb{R}^{3}$ represents the roll, pitch, and yaw angles, and $u_{i}$ ($i=1, 2, 3, 4$) represents the squared angular velocity of the $i$-th rotor. The continuous-time model of the quadrotor can be described by \citep{Raffo2010AUTO}:
\begin{equation} \label{equ_quadrotor}
	\begin{aligned}
		\ddot{\xi} &= -e_{3}g + \frac{Re_{3}}{m}U_{1},
		\\
		\ddot{\eta} &= M(\eta)^{-1}\left(\tau-C(\eta, \dot{\eta})\dot{\eta}\right),
	\end{aligned}
\end{equation}
where $e_{3} = \begin{bmatrix}
	0, 0, 1
\end{bmatrix}^{\top}$, $g = 9.81$~m/s$^2$ is the gravity acceleration, $m = 2$~kg is the quadrotor mass, $R\in \mathbb{SO}(3)$ is the rotation matrix, $M(\eta)$ is the state-dependent inertia matrix, and $C(\eta, \dot{\eta})$ is the Coriolis matrix. Detailed expressions of $R$, $M(\eta)$ and $C(\eta, \dot{\eta})$ can be found in \cite{Raffo2010AUTO}. In addition, $U_{1}$ denotes the total thrust of the rotors, and $\tau$ denotes the torques in the roll, pitch, and yaw angular directions. $U_{1}$ and $\tau$ are formulated with $u$, as follows:
\begin{equation}
	\begin{aligned}
		U_{1} &= \alpha (u_{1}+u_{2}+u_{3}+u_{4}),
		\\
		\tau &= \begin{bmatrix}
			l\alpha (-u_{2}+u_{4}) \\
			l\alpha (-u_{1}+u_{3}) \\
			\beta (-u_{1}+u_{2}-u_{3}+u_{4})
		\end{bmatrix},
	\end{aligned}
\end{equation} 
where $l=0.25$~m is the distance between the rotor and the center of mass, $\alpha=1$ is the lift constant, and $\beta=0.2$ is the drag constant. We discretize the continuous-time model \eqref{equ_quadrotor} with a sampling time of $\Delta T = 0.1$~s by using Euler's method. The control objective is to regulate the plant from the initial state $x_{0}=\begin{bmatrix}
	-1, 1, 1.5, 0, 0, 0, 0, 0, 0, 0, 0, 0
\end{bmatrix}^{\top}$ to the desired state $x_{d}=0_{12}$
%$x_{d}=\begin{bmatrix}
	%0, 0, 0, 0, 0, 0, 0, 0, 0, 0, 0, 0
	%\end{bmatrix}^{\top}$ 
by using the cloud-based MPC schemes. For the MPC formulation, the weighting matrices and vectors are selected as $Q=Q_{f}=\text{diag}\left( \begin{matrix}
	300, 300, 300, 300, 300, 300, 0, 0, 0, 0, 0, 0
\end{matrix}\right)$, $q=q_{f}=0_{12}$, $R=\text{diag}\left( \begin{matrix}
	0.1, 0.1, 0.1, 0.1
\end{matrix}\right)$ and $r=0_{4}$, and the system state and input are subjected to the constraints
$-10 \le \xi \le 10$ and $0 \le u \le 10$, respectively.
Moreover, the affine maps $\left\lbrace P_{u}, t_{u} \right\rbrace$ and $\left\lbrace P_{x}, t_{x} \right\rbrace$
	%$\left\lbrace P_{u}, t_{u} \right\rbrace$ 
are chosen as
	\[
	%{\small
		\begin{aligned}
			P_{u} &= \begin{bmatrix}
				-2 & -3 & 0 & 0 
				\\
				0.3 & -1.6 & 0 & 0 
				\\
				0 & 0 & -2 & -1
				\\
				0 & 0 & 1.2 & -3
			\end{bmatrix}, \;\;\;
			t_{u} = \begin{bmatrix}
				15 \\ -8 \\ 10 \\ -12
			\end{bmatrix},
			\\
			P_{x} &= \begin{bmatrix}
				P_{x, \xi} & 0 & 0 & 0 
				\\
				0 & P_{x, \eta} & 0 & 0 
				\\
				0 & 0 & P_{x, \dot{\xi}} & 0
				\\
				0 & 0 & 0 & P_{x, \dot{\eta}}
			\end{bmatrix}, \;\;\;
			t_{x} = \begin{bmatrix}
				t_{x, \xi} \\ t_{x, \eta} \\ t_{x, \dot{\xi}} \\ t_{x, \dot{\eta}}
			\end{bmatrix},
		\end{aligned}
		\]
		\[
		\begin{aligned}
			&P_{x, \xi} \!=\! \begin{bmatrix}
				-3 & 1.5 & 0.1 
				\\
				-1 & -2 & 0.2 
				\\
				0.5 & -1 & -2.5
			\end{bmatrix},
			t_{x, \xi} \!=\! \begin{bmatrix}
				2 \\ -3 \\ 1
			\end{bmatrix},
			P_{x, \dot{\xi}} \!=\! -2I_{3},
			t_{x, \dot{\xi}} \!=\! \begin{bmatrix}
				-1 \\ -8 \\ 7
			\end{bmatrix},
			\\
			&P_{x, \eta} \!=\! \begin{bmatrix}
				-2 \!&\! -1.5 \!&\! 0 
				\\
				-0.8 \!&\! -2.5 \!&\! 0 
				\\
				0 \!&\! -1.6 \!&\! -2
			\end{bmatrix},
			t_{x, \eta} \!=\! \begin{bmatrix}
				-5 \\ 4 \\ 6
			\end{bmatrix},
			P_{x, \dot{\eta}} \!=\! -1.5I_{3},
			t_{x, \dot{\eta}} \!=\! \begin{bmatrix}
				-3 \\ -4 \\ 3
			\end{bmatrix}.
		\end{aligned}
		%}
	\] 
	
	\begin{figure}[!t]
		\centering
		\includegraphics[width=0.48\textwidth]{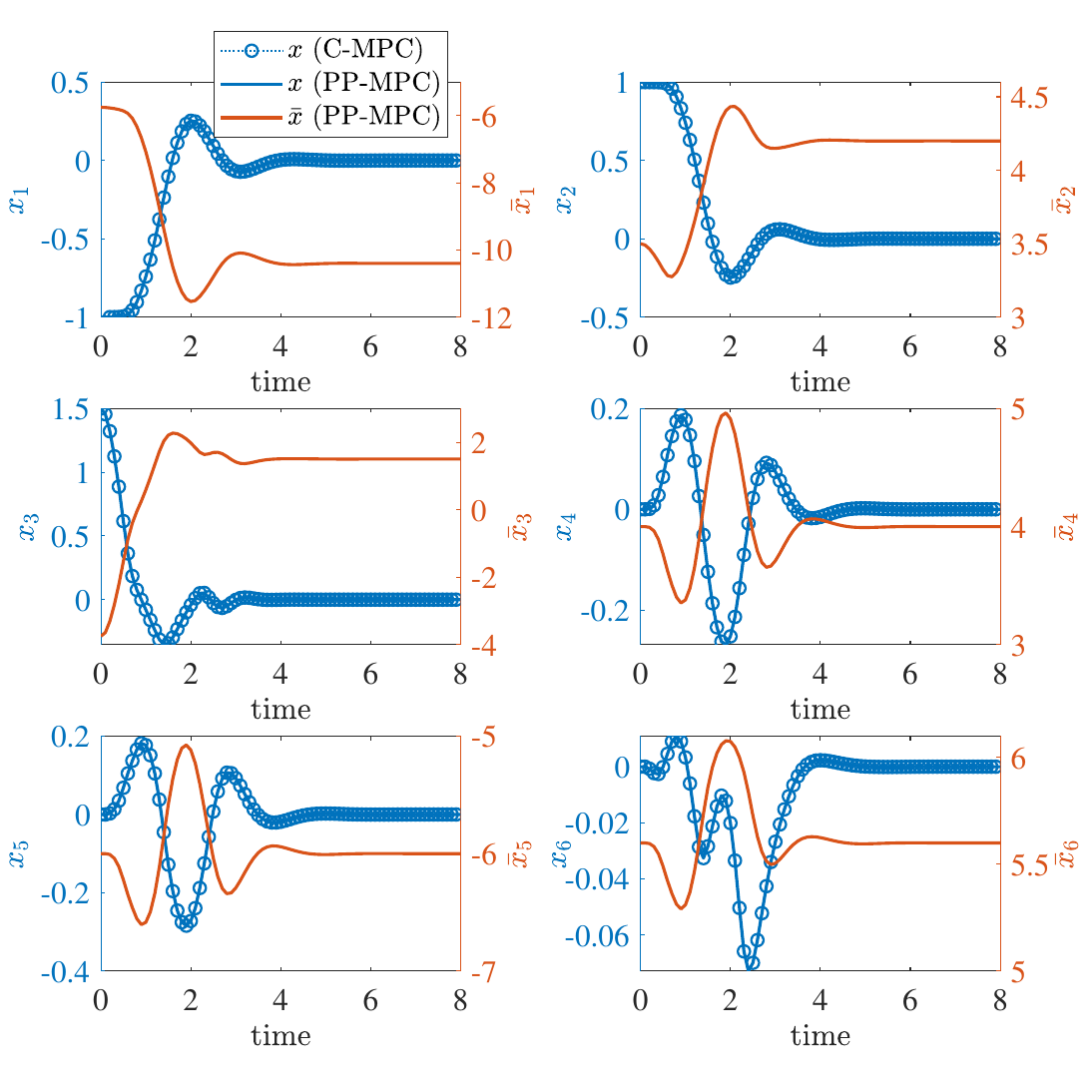}
		%\vspace{-2.5mm}
		\caption{System state evolution of conventional and privacy-preserving state-feedback MPC. C-MPC refers to conventional MPC and PP-MPC refers to privacy-preserving MPC.}
		\label{fig_x}
		%\vspace{-5mm}
	\end{figure}
	\begin{figure}[!t]
		\centering
		\includegraphics[width=0.48\textwidth]{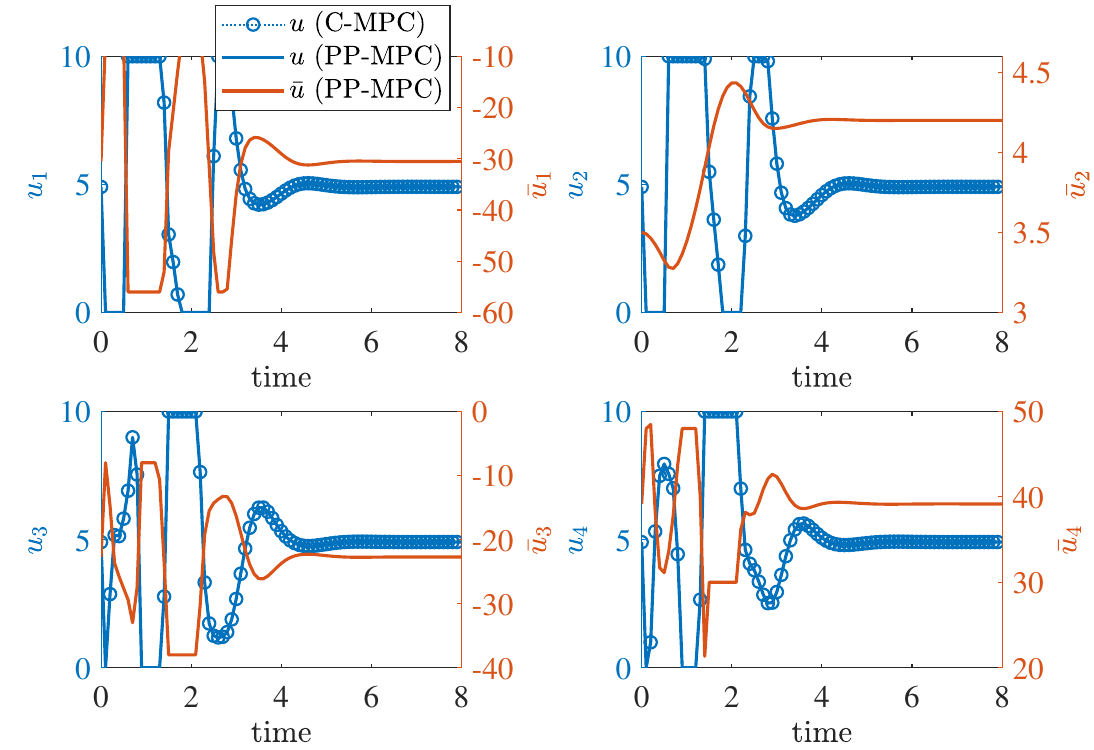}
		%\vspace{-2.5mm}
		\caption{System input evolution of conventional and privacy-preserving state-feedback MPC.}
		\label{fig_u}
		%\vspace{-5mm}
	\end{figure}
	
The state and input signals of quadrotor are privacy-sensitive, since the eavesdropper can use them to infer the quadrotor's position and velocity information and then track or attack the quadrotor. We evaluate the conventional and privacy-preserving MPC schemes with state feedback. The simulation results are presented in Figs.~\ref{fig_x} and \ref{fig_u}. Fig.~\ref{fig_x} (\ref{fig_u}) illustrates the state (input) trajectory under conventional MPC and the real and transformed state (input) trajectories under privacy-preserving MPC. It is clear that the state and input trajectories obtained from the privacy-preserving MPC are identical to the ones obtained by the conventional MPC. This aligns with the theoretical findings concluded in Lemma~\ref{Lemma:affine}, affirming that the affine transformation mechanism maintains control performance equivalent to conventional MPC. Meanwhile, as shown in Figs. \ref{fig_x} and \ref{fig_u}, under the privacy-preserving MPC, the state and input information collected by the cloud diverges significantly from the actual one. This observation underscores the efficacy of our proposed method in privacy preservation. According to Definition~\ref{def_privacy} and Theorem~\ref{theorem_privacy}, the proposed method ensures the existence of infinitely many sets of states/inputs capable of generating the same accessible information (i.e., $\bar{x}(k)$ and $\bar{u}(k)$) for the adversary. The difference among these sets could be arbitrarily large, which makes the adversary unable to infer $x(k)$ and $u(k)$.
%During the execution phase, the major difference between the conventional and privacy-preserving MPC schemes is that the latter requires the plant to conduct the affine transformation to encode $x(k)$ and decode $\bar{u}(k)$. To demonstrate the computational overhead, the computation time of the plant to implement the operations required by these two MPC schemes is recorded at each evolution step, and then the average values of the computation time are calculated. The corresponding results of the conventional and privacy-preserving MPC are $4.01\times 10^{-2}$ ms and $6.39\times 10^{-2}$ ms, respectively. This indicates that the proposed method will only result in a minor increase in computing cost.
	
\begin{figure}[!t]
	\centering
	\includegraphics[width=0.28\textwidth]{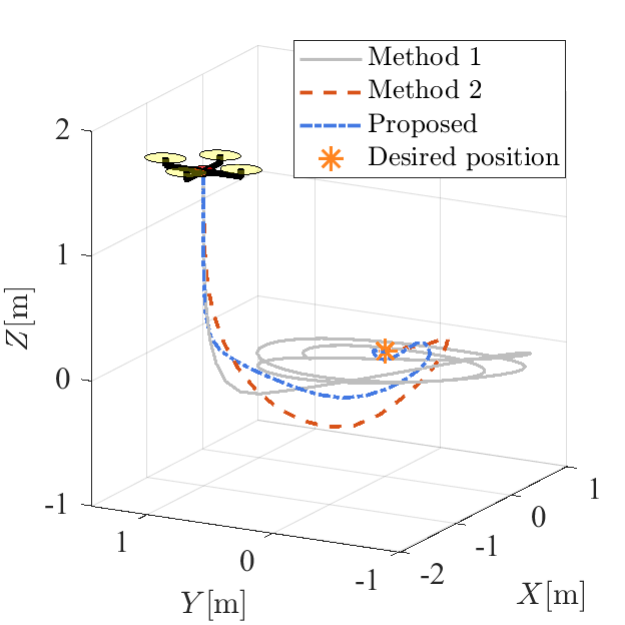}
	%\vspace{-2.5mm}
	\caption{Trajectory evolution of the quadrotor, resulting from Method 1~\citep{DarupIFAC2018}, Method 2~\citep{SultangazinTAC2021}, and the proposed method.}
	\label{fig_comparison}
	%\vspace{-5mm}
\end{figure}

For comparison, two existing privacy-preserving methods, i.e., Method 1~\citep{DarupIFAC2018} and Method 2~\citep{SultangazinTAC2021}, are tested in this simulation scenario. Method 1~\citep{DarupIFAC2018} uses homomorphic encryption to conceal sensitive information, while Method 2~\citep{SultangazinTAC2021} employs transformation-based techniques to prevent privacy leakage. Since both methods are designed for linear MPC, the nonlinear system~\eqref{equ_quadrotor} is linearized at the desired position to facilitate implementation. The motion trajectories of the quadrotor under different control schemes are illustrated in Figure~\ref{fig_comparison}. It is clear that the proposed method can effectively regulate the quadrotor to the desired position with minimal trajectory fluctuations. Moreover, Table~\ref{table_comparison} presents the accumulative cost (i.e., $\sum(x^{\top}Qx+q^{\top}x+u^{\top}Ru+r^{\top}u)$) and the average computation time required by the plant to implement the operations for different privacy-preserving methods. The proposed affine masking strategy achieves better closed-loop performance compared to Methods 1 and 2. Both the proposed strategy and Method 2 utilize similar transformation-based techniques to mask actual information, and they are more computationally efficient compared to Method 1 which relies on complicated encryption and decryption procedures.
%We run the simulation on a laptop with an Intel i7-10710U CPU with 6 cores, 1.6 GHz clock rate and 16 GB RAM.
	
\begin{table}[!t]
	\begin{center}
		\setlength{\abovecaptionskip}{3pt}
		\caption{Comparison of Accumulative Cost and Average Computation Time}\label{table_comparison}
		\scalebox{0.78}{
			\begin{threeparttable}
				\setlength{\tabcolsep}{2mm}{
					\begin{tabular}{c c c c}
						\hline
						\hline
						& Method 1 & Method 2 & Proposed \\
						\hline
						Accumulative Cost [$\times 10^{4}$] & $2.9260$ & $1.2116$ & $1.1091$
						\\
						Average Computation Time [ms] & 69.9666 & 0.0510 & 0.0499 \\
						\hline
						\hline
				\end{tabular}}
				\begin{tablenotes}
					\footnotesize
					\item[1] The average computation time refers to the time required by the plant for implementing operations under different privacy-preserving methods.
				\end{tablenotes}
			\end{threeparttable}
		}
	\end{center}
	%\vspace{-4mm}
\end{table} 

\section{Conclusion}
This paper developed an affine masking-based privacy-preserving cloud-based nonlinear MPC framework. 
We considered eavesdroppers and honest-but-curious adversaries who intend to infer the plant's system state and input and the $\infty$-diversity with unbounded diameter privacy notion was adopted. A simple yet effective affine transformation mechanism was designed to enable privacy preservation without affecting the MPC calculation.
Furthermore, the proposed method was successfully extended to output-feedback MPC. Simulation results showed that by using the proposed method, the MPC problem can be addressed without disclosing private information to the cloud. 

One thing we would like to note is that although the models are transformed in the cloud MPC implementations, one can show that the current privacy preservation scheme cannot protect the poles/zeros of the linearized system. Our future work will enhance the privacy scheme to address this issue. We will also extend this framework for systems with uncertainties (e.g., robust and stochastic MPCs), explore other metrics for privacy definition, and analyze its resilience/vulnerability to different attackers and side-knowledge. 

\bibliographystyle{elsarticle-harv}
\bibliography{IEEEfull,reference}

\begin{thebibliography}{44}
\expandafter\ifx\csname natexlab\endcsname\relax\def\natexlab#1{#1}\fi
\providecommand{\url}[1]{\texttt{#1}}
\providecommand{\href}[2]{#2}
\providecommand{\path}[1]{#1}
\providecommand{\DOIprefix}{doi:}
\providecommand{\ArXivprefix}{arXiv:}
\providecommand{\URLprefix}{URL: }
\providecommand{\Pubmedprefix}{pmid:}
\providecommand{\doi}[1]{\href{http://dx.doi.org/#1}{\path{#1}}}
\providecommand{\Pubmed}[1]{\href{pmid:#1}{\path{#1}}}
\providecommand{\bibinfo}[2]{#2}
\ifx\xfnm\relax \def\xfnm[#1]{\unskip,\space#1}\fi
%Type = Inproceedings
\bibitem[{Alexandru et~al.(2018)Alexandru, Morari and
  Pappas}]{AlexandruCDC2018}
\bibinfo{author}{Alexandru, A.B.}, \bibinfo{author}{Morari, M.},
  \bibinfo{author}{Pappas, G.J.}, \bibinfo{year}{2018}.
\newblock \bibinfo{title}{Cloud-based {MPC} with encrypted data}, in:
  \bibinfo{booktitle}{Proceedings of the {IEEE} Conference on Decision and
  Control}, pp. \bibinfo{pages}{5014--5019}.
%Type = Article
\bibitem[{Allenspach and Ducard(2021)}]{AllenspachAUTO2021}
\bibinfo{author}{Allenspach, M.}, \bibinfo{author}{Ducard, G.J.J.},
  \bibinfo{year}{2021}.
\newblock \bibinfo{title}{Nonlinear model predictive control and guidance for a
  propeller-tilting hybrid unmanned air vehicle}.
\newblock \bibinfo{journal}{Automatica} \bibinfo{volume}{132},
  \bibinfo{pages}{109790}.
%Type = Techreport
\bibitem[{Bemporad et~al.(2018)Bemporad, Bernardini, Long and Verdejo}]{GM-MPC}
\bibinfo{author}{Bemporad, A.}, \bibinfo{author}{Bernardini, D.},
  \bibinfo{author}{Long, R.}, \bibinfo{author}{Verdejo, J.},
  \bibinfo{year}{2018}.
\newblock \bibinfo{title}{Model predictive control of turbocharged gasoline
  engines for mass production}.
\newblock \bibinfo{type}{Technical Report}. SAE Technical Paper.
%Type = Article
\bibitem[{Corser et~al.(2016)Corser, Fu and Banihani}]{tracking}
\bibinfo{author}{Corser, G.P.}, \bibinfo{author}{Fu, H.},
  \bibinfo{author}{Banihani, A.}, \bibinfo{year}{2016}.
\newblock \bibinfo{title}{Evaluating location privacy in vehicular
  communications and applications}.
\newblock \bibinfo{journal}{IEEE Transactions on Intelligent Transportation
  Systems} \bibinfo{volume}{17}, \bibinfo{pages}{2658--2667}.
%Type = Inproceedings
\bibitem[{Darup and Jager(2019)}]{Darup2019CDC}
\bibinfo{author}{Darup, M.S.}, \bibinfo{author}{Jager, T.},
  \bibinfo{year}{2019}.
\newblock \bibinfo{title}{Encrypted cloud-based control using secret sharing
  with one-time pads}, in: \bibinfo{booktitle}{Proceedings of the {IEEE}
  Conference on Decision and Control}, pp. \bibinfo{pages}{7215--7221}.
%Type = Article
\bibitem[{Darup et~al.(2018a)Darup, Redder and Quevedo}]{DarupIFAC2018}
\bibinfo{author}{Darup, M.S.}, \bibinfo{author}{Redder, A.},
  \bibinfo{author}{Quevedo, D.E.}, \bibinfo{year}{2018}a.
\newblock \bibinfo{title}{Encrypted cloud-based {MPC} for linear systems with
  input constraints}.
\newblock \bibinfo{journal}{IFAC-PapersOnLine} \bibinfo{volume}{51},
  \bibinfo{pages}{535--542}.
%Type = Article
\bibitem[{Darup et~al.(2018b)Darup, Redder, Shames, Farokhi and
  Quevedo}]{SchulzeCSL2018}
\bibinfo{author}{Darup, M.S.}, \bibinfo{author}{Redder, A.},
  \bibinfo{author}{Shames, I.}, \bibinfo{author}{Farokhi, F.},
  \bibinfo{author}{Quevedo, D.E.}, \bibinfo{year}{2018}b.
\newblock \bibinfo{title}{Towards encrypted {MPC} for linear constrained
  systems}.
\newblock \bibinfo{journal}{IEEE Control Systems Letters} \bibinfo{volume}{2},
  \bibinfo{pages}{195--200}.
%Type = Inproceedings
\bibitem[{Dotzer(2005)}]{Privacy_V2V}
\bibinfo{author}{Dotzer, F.}, \bibinfo{year}{2005}.
\newblock \bibinfo{title}{Privacy issues in vehicular ad hoc networks}, in:
  \bibinfo{booktitle}{Proceedings of International Workshop on Privacy
  Enhancing Technologies}, pp. \bibinfo{pages}{197--209}.
%Type = Article
\bibitem[{{Dwork} and {Roth}(2014)}]{Dwork2014}
\bibinfo{author}{{Dwork}, C.}, \bibinfo{author}{{Roth}, A.},
  \bibinfo{year}{2014}.
\newblock \bibinfo{title}{The algorithmic foundations of differential privacy}.
\newblock \bibinfo{journal}{Foundations and Trends in Theoretical Computer
  Science} \bibinfo{volume}{9}, \bibinfo{pages}{211--407}.
%Type = Article
\bibitem[{Grossman(2009)}]{grossman2009case}
\bibinfo{author}{Grossman, R.L.}, \bibinfo{year}{2009}.
\newblock \bibinfo{title}{The case for cloud computing}.
\newblock \bibinfo{journal}{IT Professional} \bibinfo{volume}{11},
  \bibinfo{pages}{23--27}.
%Type = Inproceedings
\bibitem[{Huang et~al.(2012)Huang, Mitra and Dullerud}]{huang2012ACM}
\bibinfo{author}{Huang, Z.}, \bibinfo{author}{Mitra, S.},
  \bibinfo{author}{Dullerud, G.}, \bibinfo{year}{2012}.
\newblock \bibinfo{title}{Differentially private iterative synchronous
  consensus}, in: \bibinfo{booktitle}{Proceedings of the {ACM} Workshop on
  Privacy in the Electronic Society}, pp. \bibinfo{pages}{81--90}.
%Type = Article
\bibitem[{Hubaux et~al.(2004)Hubaux, Capkun and Luo}]{vehicle_privacy1}
\bibinfo{author}{Hubaux, J.P.}, \bibinfo{author}{Capkun, S.},
  \bibinfo{author}{Luo, J.}, \bibinfo{year}{2004}.
\newblock \bibinfo{title}{The security and privacy of smart vehicles}.
\newblock \bibinfo{journal}{{IEEE} Security $\&$ Privacy} \bibinfo{volume}{2},
  \bibinfo{pages}{49--55}.
%Type = Book
\bibitem[{Khalil(2002)}]{khalil2002nonlinear}
\bibinfo{author}{Khalil, H.K.}, \bibinfo{year}{2002}.
\newblock \bibinfo{title}{Nonlinear Systems}.
\newblock \bibinfo{publisher}{Upper Saddle River, NJ: Prentice-Hall}.
%Type = Article
\bibitem[{Li et~al.(2019)Li, Girard and Kolmanovsky}]{Li_stochastic}
\bibinfo{author}{Li, N.}, \bibinfo{author}{Girard, A.},
  \bibinfo{author}{Kolmanovsky, I.}, \bibinfo{year}{2019}.
\newblock \bibinfo{title}{Stochastic predictive control for partially
  observable markov decision processes with time-joint chance constraints and
  application to autonomous vehicle control}.
\newblock \bibinfo{journal}{Journal of Dynamic Systems, Measurement, and
  Control} \bibinfo{volume}{141}, \bibinfo{pages}{071007}.
%Type = Article
\bibitem[{Li et~al.(2023)Li, Zhang, Li, Srivastava and
  Yin}]{li2021cloudassisted}
\bibinfo{author}{Li, N.}, \bibinfo{author}{Zhang, K.}, \bibinfo{author}{Li,
  Z.}, \bibinfo{author}{Srivastava, V.}, \bibinfo{author}{Yin, X.},
  \bibinfo{year}{2023}.
\newblock \bibinfo{title}{Cloud-assisted nonlinear model predictive control for
  finite-duration tasks}.
\newblock \bibinfo{journal}{{IEEE} Transactions on Automatic Control}
  \bibinfo{volume}{68}, \bibinfo{pages}{5287--5300}.
%Type = Inproceedings
\bibitem[{{Li} et~al.(2014){Li}, {Kolmanovsky}, {Atkins}, {Lu}, {Filev} and
  {Michelini}}]{ZL_suspension}
\bibinfo{author}{{Li}, Z.}, \bibinfo{author}{{Kolmanovsky}, I.},
  \bibinfo{author}{{Atkins}, E.}, \bibinfo{author}{{Lu}, J.},
  \bibinfo{author}{{Filev}, D.}, \bibinfo{author}{{Michelini}, J.},
  \bibinfo{year}{2014}.
\newblock \bibinfo{title}{Cloud aided semi-active suspension control}, in:
  \bibinfo{booktitle}{Proceeding of the IEEE Symposium on Computational
  Intelligence in Vehicles and Transportation Systems}, pp.
  \bibinfo{pages}{76--83}.
%Type = Article
\bibitem[{{Li} et~al.(2016){Li}, {Kolmanovsky}, {Atkins}, {Lu}, {Filev} and
  {Michelini}}]{Li_Safe_Journal}
\bibinfo{author}{{Li}, Z.}, \bibinfo{author}{{Kolmanovsky}, I.},
  \bibinfo{author}{{Atkins}, E.}, \bibinfo{author}{{Lu}, J.},
  \bibinfo{author}{{Filev}, D.P.}, \bibinfo{author}{{Michelini}, J.},
  \bibinfo{year}{2016}.
\newblock \bibinfo{title}{Road risk modeling and cloud-aided safety-based route
  planning}.
\newblock \bibinfo{journal}{{IEEE} Transactions on Cybernetics}
  \bibinfo{volume}{46}, \bibinfo{pages}{2473--2483}.
%Type = Article
\bibitem[{{Li} et~al.(2017){Li}, {Kolmanovsky}, {Atkins}, {Lu}, {Filev} and
  {Bai}}]{comfort}
\bibinfo{author}{{Li}, Z.}, \bibinfo{author}{{Kolmanovsky}, I.V.},
  \bibinfo{author}{{Atkins}, E.M.}, \bibinfo{author}{{Lu}, J.},
  \bibinfo{author}{{Filev}, D.P.}, \bibinfo{author}{{Bai}, Y.},
  \bibinfo{year}{2017}.
\newblock \bibinfo{title}{Road disturbance estimation and cloud-aided
  comfort-based route planning}.
\newblock \bibinfo{journal}{{IEEE} Transactions on Cybernetics}
  \bibinfo{volume}{47}, \bibinfo{pages}{3879--3891}.
%Type = Article
\bibitem[{Liu et~al.(2016)Liu, Shi and Liu}]{LiuTIE2016}
\bibinfo{author}{Liu, M.}, \bibinfo{author}{Shi, Y.}, \bibinfo{author}{Liu,
  X.}, \bibinfo{year}{2016}.
\newblock \bibinfo{title}{Distributed {MPC} of aggregated heterogeneous
  thermostatically controlled loads in smart grid}.
\newblock \bibinfo{journal}{{IEEE} Transactions on Industrial Electronics}
  \bibinfo{volume}{63}, \bibinfo{pages}{1120--1129}.
%Type = Article
\bibitem[{{Lu} and {Zhu}(2020)}]{Lu2020}
\bibinfo{author}{{Lu}, Y.}, \bibinfo{author}{{Zhu}, M.}, \bibinfo{year}{2020}.
\newblock \bibinfo{title}{On privacy preserving data release of linear dynamic
  networks}.
\newblock \bibinfo{journal}{Automatica} \bibinfo{volume}{115},
  \bibinfo{pages}{108839}.
%Type = Article
\bibitem[{{Machanavajjhala} et~al.(2007){Machanavajjhala}, {Kifer}, {Gehrke}
  and {Venkitasubramaniam}}]{machanavajjhala2007}
\bibinfo{author}{{Machanavajjhala}, A.}, \bibinfo{author}{{Kifer}, D.},
  \bibinfo{author}{{Gehrke}, J.}, \bibinfo{author}{{Venkitasubramaniam}, M.},
  \bibinfo{year}{2007}.
\newblock \bibinfo{title}{l-diversity: Privacy beyond k-anonymity}.
\newblock \bibinfo{journal}{ACM Transactions on Knowledge Discovery from Data}
  \bibinfo{volume}{1}, \bibinfo{pages}{3--14}.
%Type = Article
\bibitem[{Mangasarian(2011)}]{Mangasarian2011}
\bibinfo{author}{Mangasarian, O.L.}, \bibinfo{year}{2011}.
\newblock \bibinfo{title}{Privacy-preserving linear programming}.
\newblock \bibinfo{journal}{Optimization Letters} \bibinfo{volume}{5},
  \bibinfo{pages}{165--172}.
%Type = Article
\bibitem[{Mayne(2014)}]{MayneAUTO2014}
\bibinfo{author}{Mayne, D.Q.}, \bibinfo{year}{2014}.
\newblock \bibinfo{title}{Model predictive control: Recent developments and
  future promise}.
\newblock \bibinfo{journal}{Automatica} \bibinfo{volume}{50},
  \bibinfo{pages}{2967--2986}.
%Type = Article
\bibitem[{McDaniel and McLaughlin(2009)}]{Mcdaniel2009}
\bibinfo{author}{McDaniel, P.}, \bibinfo{author}{McLaughlin, S.},
  \bibinfo{year}{2009}.
\newblock \bibinfo{title}{Security and privacy challenges in the smart grid}.
\newblock \bibinfo{journal}{IEEE security \& privacy} \bibinfo{volume}{7},
  \bibinfo{pages}{75--77}.
%Type = Inproceedings
\bibitem[{Munteanu et~al.(2018)Munteanu, Muradore, Merro and
  Fiorini}]{Munteanu2018}
\bibinfo{author}{Munteanu, A.}, \bibinfo{author}{Muradore, R.},
  \bibinfo{author}{Merro, M.}, \bibinfo{author}{Fiorini, P.},
  \bibinfo{year}{2018}.
\newblock \bibinfo{title}{On cyber-physical attacks in bilateral teleoperation
  systems: An experimental analysis}, in: \bibinfo{booktitle}{Proceedings of
  the {IEEE} Industrial Cyber-Physical Systems}, pp. \bibinfo{pages}{159--166}.
%Type = Inproceedings
\bibitem[{Naseri et~al.(2022)Naseri, Lucia and Youssef}]{Naseri2022ECC}
\bibinfo{author}{Naseri, A.M.}, \bibinfo{author}{Lucia, W.},
  \bibinfo{author}{Youssef, A.}, \bibinfo{year}{2022}.
\newblock \bibinfo{title}{A privacy preserving solution for cloud-enabled
  set-theoretic model predictive control}, in: \bibinfo{booktitle}{Proceedings
  of the European Control Conference}, \bibinfo{organization}{IEEE}. pp.
  \bibinfo{pages}{894--899}.
%Type = Article
\bibitem[{{National Highway Traffic Safety
  Administration}(2014)}]{NHTSA_report}
\bibinfo{author}{{National Highway Traffic Safety Administration}},
  \bibinfo{year}{2014}.
\newblock \bibinfo{title}{Vehicle-to-vehicle communications: Readiness of {V2V}
  technology for application} .
%Type = Article
\bibitem[{Nekouei et~al.(2019)Nekouei, Tanaka, Skoglund and
  Johansson}]{Nekouei2019ARC}
\bibinfo{author}{Nekouei, E.}, \bibinfo{author}{Tanaka, T.},
  \bibinfo{author}{Skoglund, M.}, \bibinfo{author}{Johansson, K.H.},
  \bibinfo{year}{2019}.
\newblock \bibinfo{title}{Information-theoretic approaches to privacy in
  estimation and control}.
\newblock \bibinfo{journal}{Annual Reviews in Control} \bibinfo{volume}{47},
  \bibinfo{pages}{412--422}.
%Type = Article
\bibitem[{Ozatay et~al.(2014)Ozatay, Onori, Wollaeger, Ozguner, Rizzoni, Filev,
  Michelini and Cairano}]{fuel_economy1}
\bibinfo{author}{Ozatay, E.}, \bibinfo{author}{Onori, S.},
  \bibinfo{author}{Wollaeger, J.}, \bibinfo{author}{Ozguner, U.},
  \bibinfo{author}{Rizzoni, G.}, \bibinfo{author}{Filev, D.},
  \bibinfo{author}{Michelini, J.}, \bibinfo{author}{Cairano, S.D.},
  \bibinfo{year}{2014}.
\newblock \bibinfo{title}{Cloud-based velocity profile optimization for
  everyday driving: A dynamic-programming-based solution}.
\newblock \bibinfo{journal}{{IEEE} Transactions on Intelligent Transportation
  Systems} \bibinfo{volume}{15}, \bibinfo{pages}{2491--2505}.
%Type = Article
\bibitem[{Petit and Shladover(2015)}]{PetitTITS2015}
\bibinfo{author}{Petit, J.}, \bibinfo{author}{Shladover, S.E.},
  \bibinfo{year}{2015}.
\newblock \bibinfo{title}{Potential cyberattacks on automated vehicles}.
\newblock \bibinfo{journal}{{IEEE} Transactions on Intelligent Transportation
  Systems} \bibinfo{volume}{16}, \bibinfo{pages}{546--556}.
%Type = Article
\bibitem[{Raffo et~al.(2010)Raffo, Ortega and Rubio}]{Raffo2010AUTO}
\bibinfo{author}{Raffo, G.V.}, \bibinfo{author}{Ortega, M.G.},
  \bibinfo{author}{Rubio, F.R.}, \bibinfo{year}{2010}.
\newblock \bibinfo{title}{An integral predictive/nonlinear {$H_{\infty}$}
  control structure for a quadrotor helicopter}.
\newblock \bibinfo{journal}{Automatica} \bibinfo{volume}{46},
  \bibinfo{pages}{29--39}.
%Type = Book
\bibitem[{Rawlings et~al.(2017)Rawlings, Mayne and Diehl}]{Rawlings2017MPC}
\bibinfo{author}{Rawlings, J.B.}, \bibinfo{author}{Mayne, D.Q.},
  \bibinfo{author}{Diehl, M.}, \bibinfo{year}{2017}.
\newblock \bibinfo{title}{Model Predictive Control: Theory, Computation, and
  Design}.
\newblock \bibinfo{publisher}{Nob Hill Publishing}.
%Type = Article
\bibitem[{{Sankar} et~al.(2013){Sankar}, {Rajagopalan} and {Poor}}]{Sankar2013}
\bibinfo{author}{{Sankar}, L.}, \bibinfo{author}{{Rajagopalan}, S.R.},
  \bibinfo{author}{{Poor}, H.V.}, \bibinfo{year}{2013}.
\newblock \bibinfo{title}{Utility-privacy tradeoffs in databases: {A}n
  information-theoretic approach}.
\newblock \bibinfo{journal}{{IEEE} Transactions on Information Forensics and
  Security} \bibinfo{volume}{8}, \bibinfo{pages}{838--852}.
%Type = Inproceedings
\bibitem[{Schlor et~al.(2021)Schlor, Hertneck, Wildhagen and
  Allg{\"o}wer}]{Schlor2021CDC}
\bibinfo{author}{Schlor, S.}, \bibinfo{author}{Hertneck, M.},
  \bibinfo{author}{Wildhagen, S.}, \bibinfo{author}{Allg{\"o}wer, F.},
  \bibinfo{year}{2021}.
\newblock \bibinfo{title}{Multi-party computation enables secure polynomial
  control based solely on secret-sharing}, in: \bibinfo{booktitle}{Proceedings
  of the {IEEE} Conference on Decision and Control}, pp.
  \bibinfo{pages}{4882--4887}.
%Type = Inproceedings
\bibitem[{Schl{\"u}ter and Darup(2020)}]{NilsCDC2020}
\bibinfo{author}{Schl{\"u}ter, N.}, \bibinfo{author}{Darup, M.S.},
  \bibinfo{year}{2020}.
\newblock \bibinfo{title}{Encrypted explicit {MPC} based on two-party
  computation and convex controller decomposition}, in:
  \bibinfo{booktitle}{Proceedings of the IEEE Conference on Decision and
  Control}, pp. \bibinfo{pages}{5469--5476}.
%Type = Article
\bibitem[{Shamir(1979)}]{Shamir1979}
\bibinfo{author}{Shamir, A.}, \bibinfo{year}{1979}.
\newblock \bibinfo{title}{How to share a secret}.
\newblock \bibinfo{journal}{Communications of the ACM} \bibinfo{volume}{22},
  \bibinfo{pages}{612--613}.
%Type = Article
\bibitem[{Sultangazin and Tabuada(2021)}]{SultangazinTAC2021}
\bibinfo{author}{Sultangazin, A.}, \bibinfo{author}{Tabuada, P.},
  \bibinfo{year}{2021}.
\newblock \bibinfo{title}{Symmetries and isomorphisms for privacy in control
  over the cloud}.
\newblock \bibinfo{journal}{{IEEE} Transactions on Automatic Control}
  \bibinfo{volume}{66}, \bibinfo{pages}{538--549}.
%Type = Inproceedings
\bibitem[{Wang et~al.(2011)Wang, Ren and Wang}]{WangINFOCOM2011}
\bibinfo{author}{Wang, C.}, \bibinfo{author}{Ren, K.}, \bibinfo{author}{Wang,
  J.}, \bibinfo{year}{2011}.
\newblock \bibinfo{title}{Secure and practical outsourcing of linear
  programming in cloud computing}, in: \bibinfo{booktitle}{Proceedings of
  {IEEE} {INFOCOM}}, pp. \bibinfo{pages}{820--828}.
%Type = Inproceedings
\bibitem[{Weeraddana et~al.(2013)Weeraddana, Athanasiou, Fischione and
  Baras}]{Weeraddana2013CDC}
\bibinfo{author}{Weeraddana, P.C.}, \bibinfo{author}{Athanasiou, G.},
  \bibinfo{author}{Fischione, C.}, \bibinfo{author}{Baras, J.S.},
  \bibinfo{year}{2013}.
\newblock \bibinfo{title}{Per-se privacy preserving solution methods based on
  optimization}, in: \bibinfo{booktitle}{Proceedings of the {IEEE} Conference
  on Decision and Control}, pp. \bibinfo{pages}{206--211}.
%Type = Article
\bibitem[{Weeraddana and Fischione(2017)}]{Weeraddana2017IFAC}
\bibinfo{author}{Weeraddana, P.C.}, \bibinfo{author}{Fischione, C.},
  \bibinfo{year}{2017}.
\newblock \bibinfo{title}{On the privacy of optimization}.
\newblock \bibinfo{journal}{IFAC-PapersOnLine} \bibinfo{volume}{50},
  \bibinfo{pages}{9502--9508}.
%Type = Article
\bibitem[{Xu et~al.(2021)Xu, Deng, Zhang, Qiu and Bao}]{Xu2021IS}
\bibinfo{author}{Xu, Y.}, \bibinfo{author}{Deng, G.}, \bibinfo{author}{Zhang,
  T.}, \bibinfo{author}{Qiu, H.}, \bibinfo{author}{Bao, Y.},
  \bibinfo{year}{2021}.
\newblock \bibinfo{title}{Novel denial-of-service attacks against cloud-based
  multi-robot systems}.
\newblock \bibinfo{journal}{Information Sciences} \bibinfo{volume}{576},
  \bibinfo{pages}{329--344}.
%Type = Inproceedings
\bibitem[{Xu and Zhu(2015)}]{Xu2015ACM}
\bibinfo{author}{Xu, Z.}, \bibinfo{author}{Zhu, Q.}, \bibinfo{year}{2015}.
\newblock \bibinfo{title}{Secure and resilient control design for cloud enabled
  networked control systems}, in: \bibinfo{booktitle}{Proceedings of the first
  {ACM} workshop on cyber-physical systems-security and/or privacy}, pp.
  \bibinfo{pages}{31--42}.
%Type = Inproceedings
\bibitem[{Xu and Zhu(2017)}]{Xu2017}
\bibinfo{author}{Xu, Z.}, \bibinfo{author}{Zhu, Q.}, \bibinfo{year}{2017}.
\newblock \bibinfo{title}{Secure and practical output feedback control for
  cloud-enabled cyber-physical systems}, in: \bibinfo{booktitle}{Proceedings of
  the {IEEE} Conference on Communications and Network Security}, pp.
  \bibinfo{pages}{416--420}.
%Type = Article
\bibitem[{Xue et~al.(2014)Xue, Wang and Roy}]{vehicle_privacy}
\bibinfo{author}{Xue, M.}, \bibinfo{author}{Wang, W.}, \bibinfo{author}{Roy,
  S.}, \bibinfo{year}{2014}.
\newblock \bibinfo{title}{Security concepts for the dynamics of autonomous
  vehicle networks}.
\newblock \bibinfo{journal}{Automatica} \bibinfo{volume}{50},
  \bibinfo{pages}{852--857}.

\end{thebibliography}

\end{document}